# Deep learning in medical image registration: introduction and survey

Ahmad Hammoudeh, Stéphane Dupont
MAIA Artificial Intelligence Lab

## Abstract

Image registration (IR) is a process that deforms images to align them with respect to a reference space, making it easier for medical practitioners to examine various medical images in a standardized reference frame, such as having the same rotation and scale. This document introduces image registration using a simple numeric example. It provides a definition of image registration and a constraint-based analysis, along with a space-oriented symbolic representation. This review covers various aspects of image transformations, including affine, deformable, invertible, and bidirectional transformations, as well as medical image registration algorithms such as Voxelmorph, Demons, SyN, Iterative Closest Point, and SynthMorph. It also explores atlas-based registration and multistage image registration techniques, including coarse-fine and pyramid approaches. Furthermore, this survey paper discusses medical image registration taxonomies, datasets, evaluation measures, such as correlation-based metrics, segmentation-based metrics, processing time, and model size. It also elaborates on IR applications in image-guided surgery, motion tracking, and tumor diagnosis. Finally, the document addresses future research directions, including the further development of transformers.

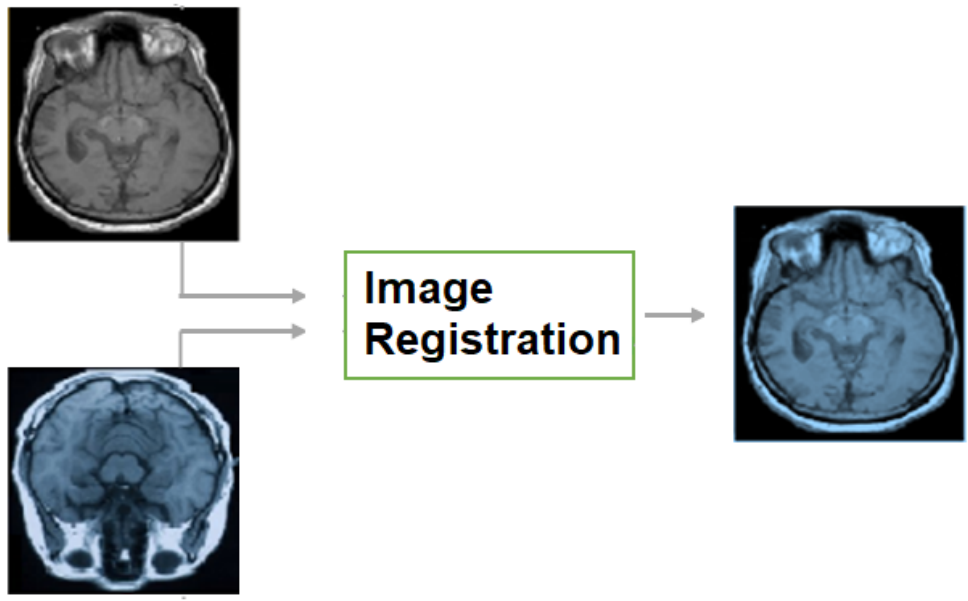

Deep learning in medical image registration: introduction and survey

Contents

# Deep learning in medical image registration: introduction and survey

## Nomenclature

| | |
|---|---|
| $b_x$ | translation on the X-axis |
| $b_y$ | translation on the Y-axis |
| $b_z$ | translation on the Z-axis |
| $C_{ij}^{pq}$ | correspondence of image p in space i and image q in space j |
| DT | average training time |
| E | the total number of elements in a set |
| e (as in xe) | stands for the order of an element in a set $X_{\emptyset i}^p = \{x1_{\emptyset i}^p, x2_{\emptyset i}^p, x3_{\emptyset i}^p, \ldots\ldots xE_{\emptyset i}^p\}$, $Y_{\emptyset i}^p = \{y1_{\emptyset i}^p, y2_{\emptyset i}^p, y3_{\emptyset i}^p, \ldots\ldots yE_{\emptyset i}^p\}$ |
| $F_{ij}^p$ | a mapping between $X_{ij}^p$ & $Y_{ij}^p$ such that $Y_{ij}^p = F_{ij}^p(X_{ij}^p)$ |
| $I_{\emptyset i}^p$ (or $I_i^p$) | $<X_{\emptyset i}^p, Y_{\emptyset i}^p>$ an image P that connects $X_{\emptyset i}^p$ and $Y_{\emptyset i}^p$ |
| $L_{\emptyset i}^p$ | labels associated with $I_{\emptyset i}^p$ that describe properties of elements in $I_{\emptyset i}^p$ such as a segmentation category |
| Lp | the number of points in image P |
| $M_{\emptyset i}^p$ | a set of marks that includes selected elements or points of image P such that $M_{\emptyset i}^p \in I_{\emptyset i}^p$ |
| Mp | the number of landmarks in image P |
| N | the number of examples/samples in a dataset |
| O | an objective function that yields a smaller value when the registration is closer to the desired. For example, $O = \sum_{p=1}^{N}(Y_{\emptyset j}^p - Y_{ij}^p)^2$ which measures the square difference between a registered codomain $Y_{ij}^p$ and a ground truth $Y_{\emptyset j}^p$ |
| RT | average registration runtime |
| $T_{ij}$ | a domain mapping between $X_{\emptyset i}$ & $X_{\emptyset j}$ |
| $t_{r,start}^p$ | the time at which image p is loaded to an IR model |
| $t_{r,end}^p$ | the time at which image p is registered |
| $\|v\|$ | length of vector v |
| $X_{\emptyset i}^p$ (or $X_i^p$) | domain values of an image p in space i, where $\emptyset$ is a reference unknown codomain (used with raw data). p is an index of a registration example in a dataset. |
| $X_{ij}$ | the transformed domain after applying $T_{ij}$ to $X_{\emptyset i}$ such that $X_{ij} = T_{ij}(X_{\emptyset i})$ |
| $X_{i-j}$ | the outcome of applying $T_{i-j}$ to $X_{\emptyset i}$ such that $X_{i-j} = T_{i-j}(X_{\emptyset i})$ but before any postprocessing like resampling |
| xe | element number e in a set X, $X_{\emptyset i}^p = \{x1_{\emptyset i}^p, x2_{\emptyset i}^p, x3_{\emptyset i}^p, \ldots\ldots xE_{\emptyset i}^p\}$, |
| $Y_{\emptyset i}^p$ (or $Y_i^p$) | codomain values of an image p in space i, where $\emptyset$ is a reference unknown codomain (used with raw data). p is an index of a registration example in a dataset. |
| $Y_{i-j}$ | codomain values after a transformation $T_{i-j}$ but before any post-processing (e.g., interpolation) |
| ye | element number e in a set Y, $Y_{\emptyset i}^p = \{y1_{\emptyset i}^p, y2_{\emptyset i}^p, y3_{\emptyset i}^p, \ldots\ldots yE_{\emptyset i}^p\}$ |
| $\Delta_{ij}^p$ | a displacement field that transforms image p from space i to j. |
| $\varphi_{ij}^p$ | a registration field that transforms image p from space i to j. |
| θ | rotation angle |
| ′ | the apostrophe indicates ground truth, for example, $<X'^p_{ij}, Y'^p_{ij}>$ is the ground truth outcome <domain and codomain> of image p after IR to space j. |
| ~ | the ~ sign indicates a post-transformation outcome. |
| # | number of |
| σ | standard deviation |
| ∩ | intersection |
| ∪ | union |

Abbreviations

| | |
|---|---|
| 2D | two dimensional |
| 3D | three dimensional |
| AI | artificial intelligence |
| CC | cross-correlation |
| COM | center of mass |
| CMYK | a color system in which cyan, magenta, yellow, and black are the basic colors |
| CT | computerized tomography |
| CNNs | convolutional neural networks |
| Dist | distance measure |
| DL | deep learning |
| DSC | dice score |
| e.g. | for example |
| F1 | F score |
| FN | false negative |
| FP | false positive |
| GANs | generative adversarial networks |
| HD | Hough distance |
| IR | image registration |
| Inf | infimum |
| i.e. | that is |
| J | Jacobian |
| JOCA | the determinant of Jacobian |
| MIR | medical image registration |
| ML | machine learning |
| MR | magnetic resonance imaging |
| MSE | mean square error |
| N | no |
| nCC | normalized cross-correlation |
| nLCC | normalized local cross-correlation |
| NN | neural networks |
| PET | positron emission tomography |
| Prox | a proximity measure |
| RGB | a color system in which red, green, and blue are the basic colors |
| RL | reinforcement learning |
| RMSE | root mean square error |
| ROI | region of interest |
| SDlogJ | standard deviation of log Jacobian |
| Sup | supremum |
| TRE | target registration error |
| TN | true negative |
| TP | true positive |
| w/o | without |
| US | ultra-sound |
| Y | yes |

# Deep learning in medical image registration: introduction and survey

## 1.0 Introduction

### 1.1 Image registration etymology: in dictionaries

When a novice human reads or hears the concept of "image registration" for the first time, the word "registration" may not provide a clue about what image registration engineers do. A curious non-native English speaker may look for a hint in a dictionary such as Oxford or Cambridge, but none of the listed senses in the dictionaries seem related. In the dictionary, the word "registration" is mainly associated with an entry in an official record or list, such as the addition of a new citizen to a national register or the enrollment of a student in a course (an entry in the record of enrolled students). Similarly, the license plate number on the back and front of a car is called a registration number in British English (an entry in the record of licensed cars). A related sense is found in Merriam-Webster's dictionary under the word "register," but not under the word "registration," which defines "register" (noun) as a correct alignment. Nevertheless, the Cambridge and Oxford dictionaries do not list a related sense under "register" or "registration." Links are in Table 1.

Table 1. Registration in dictionaries (Oxford, Cambridge, Meriam Webster)

| Dictionary | Word | Link |
|---|---|---|
| Oxford | Registration | https://www.oxfordlearnersdictionaries.com/definition/english/registration?q=registration |
| Oxford | Register (noun) | https://www.oxfordlearnersdictionaries.com/definition/english/register_2 |
| Oxford | Register (verb) | https://www.oxfordlearnersdictionaries.com/definition/english/register_1?q=register |
| Cambridge | Registration | https://dictionary.cambridge.org/dictionary/english/registration |
| Cambridge | Register | https://dictionary.cambridge.org/dictionary/english/register |
| Meriam webster | Registration | https://www.merriam-webster.com/dictionary/registration |
| Meriam webster | Register | https://www.merriam-webster.com/dictionary/register |

### 1.2 Image registration etymology: in the printing industry

The concept of 'image registration' in computer vision could have been coined under the influence of the early printing industry. In the printing industry, registration is the process of getting an image printed at the same location on the paper each time. It also means the perfect alignment of printing components (e.g., dots, lines, colors) with respect to each other. Figure 1 shows an example of printing misalignments. The misalignment in early printing machines depended on the initial settings in addition to the movement of a paper while it runs through the printing machine. Hence, marks like crosshairs were used to be printed on paper boundaries to check a popper alignment/registration (Stallings, 2010). You may have seen a crosshair like the one shown on the right of Figure 1 in old printed documents. In addition to ink printing on papers, registration covered other kinds of printing such as embossing and metallic foiling.

In color printing, basic colors were printed one color after another. For example, the basic colors of the CMYK color model are cyan, magenta, yellow, and key/black, which form the acronym of CMYK. A misalignment between colors may result in overlapping replicas (Wikipedia 2023).

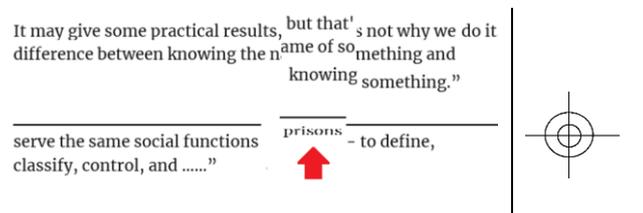

Figure 1. An example of printing misalignment and a crosshair on the right boundary. Crosshairs were used in the early printing industry to check misalignments like the misalignment between copies/pages

In summary, the concept of image registration in computer vision seems to have been influenced by the printing industry. In computer vision, it is common to call a computer image that will be aligned a "moving image" and the reference image a "fixed image". The naming of a "moving" and a "fixed" image suit the movement of a paper in a printing process, however, it is still very common to read "moving image" and "fixed image" in computer image registration papers, despite the lack of a moving part as IR is done digitally by computer algorithms only. The sense of registration in the printing industry can be seen as a narrow case of an expanding IR arena. IR includes aligning identical images, aligning images of non-rigid objects, and aligning images of different objects and/or different dimensions (like aligning a 2D X-ray image with a 3D MRI image).

Deep learning in medical image registration: introduction and survey

In the previous paragraphs, a potential connection was drawn between the concept of IR and the registration printing industry. Another potential, but less obvious, connection is between IR and registration in the music industry. Registration is known to organists (musicians who play organ) as the selection of organ stops. An organ stop is a part of an organ that controls the flow of the air to certain pipes, hence basically a musician combines stops to generate sounds. What is common between organ registration and IR is that both are processes of finding a configuration for an intended outcome. That outcome is a melody in the case of organ registration and an aligned image in the case of IR. Can organ registration be considered a type of IR? Answering such questions entails a full technical definition of image registration (see the next section).

**1.3 Image registration definition**
Humans align objects mentally before deciding whether two rotated objects are similar or not according to cognitive psychology (Cooper, 1975). Likewise, it is easier for medical practitioners to compare aligned medical images. To demonstrate this, a reader can compare the left side and the right side of Figure 2.

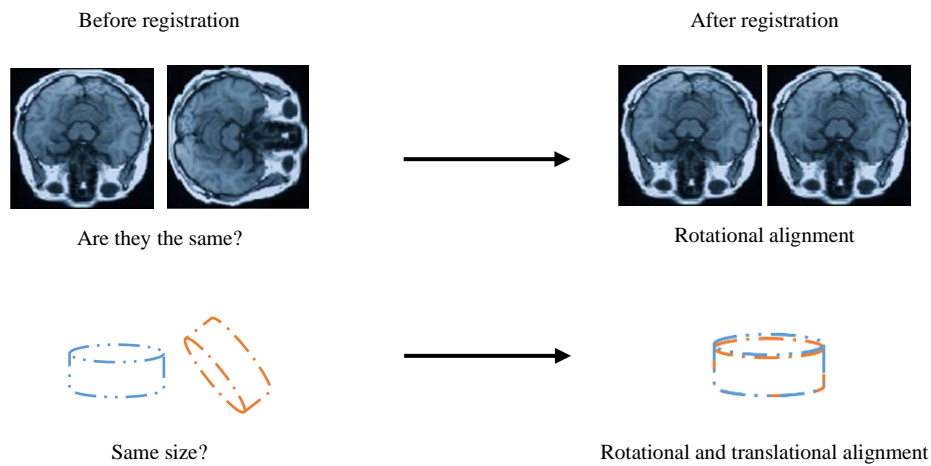

Figure 2. Examples of mental alignment

*1.3.1 IR definitions in the literature*
Image registration definitions in the literature of medical image registration can be categorized into three main definitions:

> *Definition 1*: Finding a transformation between **two images** that are related to each other such that the images are of the same object or similar objects, the same region, or similar regions (Stewart et al., 2004).
> *Definition 2*: The process of overlaying two or more images of the **same scene** taken at different times, from different viewpoints, and/or by different sensors (Zitova et al., 2003).
> *Definition 3*: The process of transforming different images into **one coordinate system** with matched content (Chen, X., Wang, et al, 2022).

The first definition limited registration to two images. The second definition extended the number of images beyond two, but it limited registration to the "same scene" while the first definition included "similar scenes". Hence, the alignment of an image of a human brain and another of a monkey's brain is covered by the first definition, but not clearly by the second. For medical images, the alignment between images of different patients is known as inter-patient registration. Hence 'inter-patient' registration is not covered by the second definition. The first two definitions limited registration to "same" or "similar" scenes, but no clear definition was provided to distinguish what can be considered similar and what cannot. Similarity in physics entails a preservation of at least one quantity such that the quantity measured in case 1 (first image) will be equal to that in case 2 (second image). Given the fundamental definition of similarity, we argue that the concept of "similarity" should not be part of the definition unless there is really a preserved quantity determined. Some may argue that the preserved quantity is the loss function used for optimization. Although that optimization function guides an approximate solution for a registration problem, that function is not grounded (the optimization function itself can be adapted). Moreover, that optimization function or similarity measure is not preserved for the images being registered. Hence, we argue that the concept of similarity should not be part of a definition to IR, but it is a part of an engineered IR solution.

# Deep learning in medical image registration: introduction and survey

The proposed definition of image registration which will be introduced later in this section eliminates the concept of similarity, instead, it considers that any images with a correspondence relation can be registered, for example, the registration of an image of an orange with another of an apple is possible. Likewise the registration of a sound with a video.

The third definition introduced a "coordinate system", but does image registration entail a coordinate system? If yes, is it one coordinate system or more? The answer will be no if a single example of a registration process can be done without a coordinate system (falsifiability). Let's assume a mapping between two rectangles is expressed by colors where the corresponding points share the same color. The need for a coordinate system in image registration is similar to the need for an absolute reference frame in Newtonian physics. In Newtonian physics, measurements of moving objects are recorded with respect to an absolute reference point that is chosen randomly and called the origin of a coordinate system. However, that coordinate system is a just tool that is not an essential part of the world. In his theory of relativity, Einstein suggested a relative frame of reference where the movement of an object is measured with reference to any point. An absolute coordinate system is a useful tool that makes solving the IR problem easier, especially to define locations, but that tool is not essential since we can define an object without it using a graph or a set. Our argument here is not to propose the elimination of the coordinate systems (yet), but to propose a more accurate concept than the coordinate system, which is the concept of "space". The alignment of images happens in a "space". That space may or may not have a coordinate system associated with it. The presence or absence of a coordinate system does not affect the components of a space. Some spaces do not even accept a coordinate system such as topological spaces and spaces of infinite dimensions. Replacing the concept of the "coordinate system" by the concept of the "space" does not make the definition more accurate only, but also proposes solving the registration problem in other mathematical spaces beyond the coordinate system.

Table 2 shows a list of image registration definitions extracted from review papers on MIR along with the constraints that each definition imposed. The differences between IR definitions are about imposing the following constraints:

> C1 registration is limited to two images,
> C2 registration is limited to the "same scene",
> C3 a coordinate system is essential.

Table 2: Image registration definitions in MIR review papers

| Paper | Definition | Constraint |
|---|---|---|
| (Decuyper et al., 2021) | "the process of aligning two images so that anatomical features would spatially coincide. This is required when analyzing pairs of images that were taken at different times or taken by different imaging modalities." | C1 |
| (Zitova et al., 2003) | "Image registration is the process of overlaying images (two or more) of the same scene taken at different times, from different viewpoints, and/or by different sensors. The registration geometrically aligns two images (the reference and sensed images)." | C2 |
| (Chen, X., Wang et al, 2022) | "the process of aligning two or more images into one coordinate system with matched contents." | C3 |
| (Haskins et al., 2020) | "The process of transforming different image datasets into one coordinate system with matched imaging contents, which has significant applications in medicine. Registration may be necessary when analyzing a pair of images that were acquired from different viewpoints, at different times, or using different sensors/modalities" | C3 |
| (Chen, X. et al 2021) | "The process of identifying a spatial transformation that maps two (pair-wise registration) or more (group-wise registration) images to a common coordinate frame such that corresponding anatomical structures are optimally aligned, or in other words, a voxel-wise 'correspondence' is established between the images." | C3 |
| (Abbasi et al., 2022) | "the proper procedure for transforming different images into one coordinate system. In this process, we align multiple images of the same object into a single cumulative amalgamated image, taking into account the rotation, skew, and scale of each separate image. This is what is called image registration, known as image fusion or image matching" | C2, C3 |
| (Talwar et al., 2021) | To "align an image to a reference image" | C1 |

*1.3.2 IR definition*

The definition considered in this document does not impose any of the constraints C1, C2, or C3. Briefly, Image registration is an alignment of images. But we still need to define alignment and images.

The **alignment** is finding a space k (not necessarily a coordinate system) in which a correspondence relation is satisfied such that correspondent elements are in proximity. For instance, given an image $I_{\emptyset i}^{p} = <X_{\emptyset i}^{p}, Y_{\emptyset i}^{p}>$, and $I_{\emptyset j}^{q} = <X_{\emptyset j}^{q}, Y_{\emptyset j}^{q}>$ with a correspondence set between them $C_{ij}^{pq} = \{(xe_{\emptyset i}^{p}, ye_{\emptyset i}^{p}) \leftrightarrow (xe_{\emptyset j}^{q}, ye_{\emptyset j}^{q})\}$, the images p and q are registered if there is a space k in which the correspondent points are almost equal in that space. That can be expressed as in Equation 1.

$$T_{ik}(xe^p_{\emptyset i}, ye^p_{\emptyset i}) \approx T_{jk}(xe^q_{\emptyset j}, ye^q_{\emptyset j}) \text{ for } \forall \text{ correspondence points} \in C^{pq}_{ij} \quad (1)$$

where
- $(xe^p_{\emptyset i}, ye^p_{\emptyset i}) \leftrightarrow (xe^q_{\emptyset j}, ye^q_{\emptyset j})$ are correspondent points between image p in space i, and image q in space j,

- $T_{ik}, T_{jk}$ transformations map spaces i, and j (respectively) to space k.

An **image** is a representation of a function that maps a space/set X called domain to a space/set Y called co-domain. This definition may seem a generic definition that goes beyond digital graphics in the sense that a mathematical function y=$x^2$ can be considered an image under this definition, which is true. The discussion of what is an image and what is not is beyond the scope of this document. However, interested readers are referred to (Mitchel, 1984) who provided an interesting discussion based on Wittgenstein's philosophy, in which Mitchell thinks of a family of images that includes graphical, optical, mental, and verbal images. This definition aligns with the classical concept of a digital image. A digital graphical image, which is a discrete function, can be thought of as a set of samples (e.g., recorded by a sensor) from a continuous function. That continuous function is a scene/object/manifold in the world. However, a sensor is no longer essential to acquire a digital graphical image since digital graphical images can be created virtually, using graphics tools or deep learning (GANs for example).

Overall, the definition above is distinguished from the image registration definitions mentioned earlier in the following points:

1- Registration is not limited to two images, instead multiple images can be registered in a space called correspondence space.
2- The possibility of registration is not limited to "similar" images, instead, registration is between any images with a correspondence relation (as in mathematics). The correspondence relation can be explicit as in the iterative closest point method, or implicit as in deep learning approaches. Both approaches are explained in section 7.3 titled MIR algorithms.
3- A coordinate system is not part of the definition; however, the concept of space (as in mathematics) is used instead for a more accurate and generic definition.

Although constraints C1, C2, or C3 were common in the surveyed MIR papers between 2021 and 2022, some IR definitions that do not impose any of those constraints were found in earlier publications. (Modersitzki, 2003) defined IR as finding an *optimal* geometric transformation between *corresponding* images, where the notions '*optimal*' and '*corresponding*' were considered dependent on the application. Unfortunately, just the first chapter of (Modersitzki, 2003)'s book was found open access. In (Brown, 1992), two definitions were provided. while the first definition in the abstract "to match two or more pictures taken, for example, at different times, from different sensors, or different. viewpoints." does not explicitly impose any of the constraints C1-C3, the second definition in a subsection titled "definition" imposed C1: "a mapping between two images both spatially and with respect to intensity." (Fitzpatrick et al., 2000) defined IR as "the determination of a geometrical transformation that aligns points in one view of an object with corresponding points in another view of that object or another object.", which seems to comply with C1, but rejects C2, and doesn't impose C3.

## 2.0 Introductory example
Readers who have some knowledge of image processing or modern algebra are advised to skip this section which targets novice readers. This section demonstrates an image transformation pipeline using a simple example (see Figure 3).

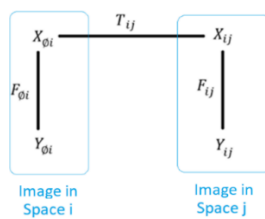

Figure 3. A transformation of an image from space i to space j

Deep learning in medical image registration: introduction and survey

## 2.1 Image representations

Let P be a 2D digital image of 2x2 pixels in a Euclidean space as shown in Figure 4. The domain values $X_{\emptyset i}^p$ can be represented in a set of tuples {(m, n)} where m, n are integers, $0 \leq m < 2$, $0 \leq n < 2$. Explicitly $X_{\emptyset i}^p$ = {(0,0), (0,1), (1,0), (1,1)}, where the first number in the tuple is the row and the second is the column in which a pixel is located such that counting starts from the top left corner of the image. The codomain values $Y_{\emptyset i}^p$ is a set of 2x2 = 4 items, each item represents a color. $Y_{\emptyset i}^p$ = {(a), (b), (c), (d)}. Colors in this example were represented as symbols for simplicity. There are multiple other ways to represent the domain/codomain values. For example, the codomain values of an RGB image consist of 3 numbers that represent the intensities of the basic colors (red, green, blue), which if mixed yield the color of the pixel.

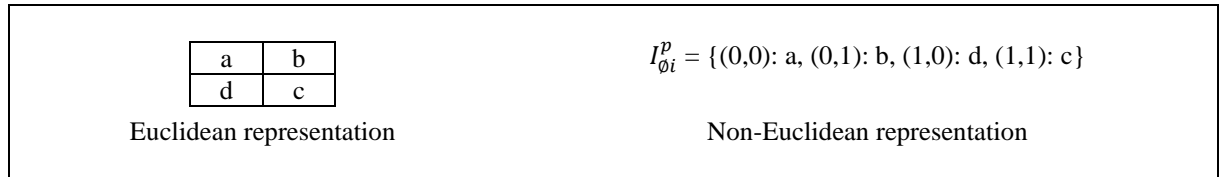

Figure 4. Two representations of an image: Euclidean (left), and non-Euclidean (right).

$F_{\emptyset i}^p$ is a mapping between $X_{\emptyset i}^p$ & $Y_{\emptyset i}^p$ like how the blue lines in Figure 5 connect elements in the left rectangle (domain values) to elements in the right rectangle (co-domain values). $F_{\emptyset i}^p$ can be represented, in some other cases, by an algebraic expression between X and Y.

## 2.2 Image transformation

A toy example of image transformation is represented in Figure 5. T$_{ij}$ is a domain transformation that relocates pixels 1 unit in a counterclockwise rotation. T$_{ij}$ shown in Figure 5B replaced the domain values with new ones. For example, the domain value (0,0) in Figure 5A became (1,1) in Figure 5C.

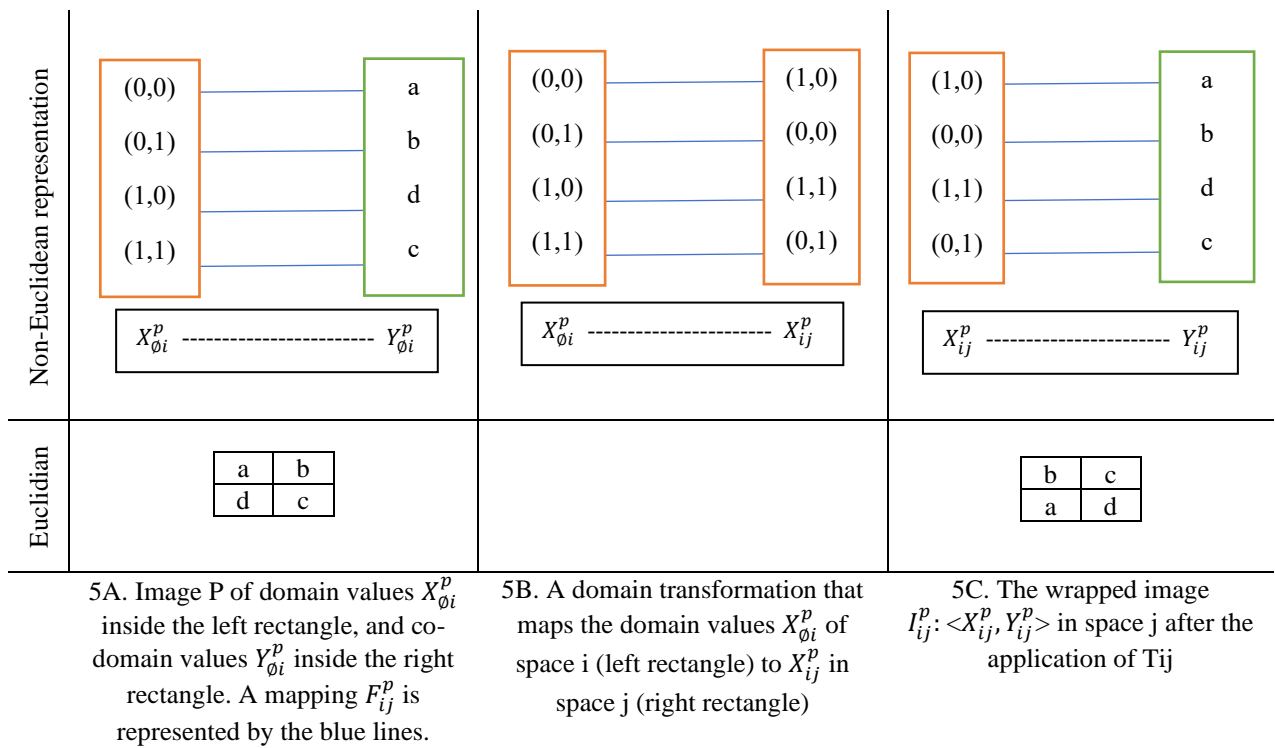

5A. Image P of domain values $X_{\emptyset i}^p$ inside the left rectangle, and co-domain values $Y_{\emptyset i}^p$ inside the right rectangle. A mapping $F_{ij}^p$ is represented by the blue lines.

5B. A domain transformation that maps the domain values $X_{\emptyset i}^p$ of space i (left rectangle) to $X_{ij}^p$ in space j (right rectangle)

5C. The wrapped image $I_{ij}^p$: <$X_{ij}^p, Y_{ij}^p$> in space j after the application of Tij

Figure 5. An image transformation

Deep learning in medical image registration: introduction and survey

*Image deformation using a displacement field:*

A displacement field ΔX represents domain relocation distances (see Equation 2), such that a 2D displacement value of <1,-1> moves a pixel 1 unit on the horizontal axis and -1 unit on the vertical axis.

$$X^p_{i-j} = T_{i-j}(X^p_{\emptyset i}) = X^p_{\emptyset i} + \Delta X \tag{2}$$

Figure 6 shows a displacement field ΔX estimated by an algorithm given a pair of fixed and moving images. The moving image $I^p_{\emptyset i}$ = {'0,0': a, '0,1': b, '1,0': d, '1,1': c}. The displacement field estimated by a registration algorithm is ΔX = {'0,0': <1,0>, '0,1': <0,-1>, '1,0': <0,2>, '1,1': <-1,0>}. The goal is to obtain a transformed image $I^p_{ij}$ by deforming the moving image $I^p_{\emptyset i}$.

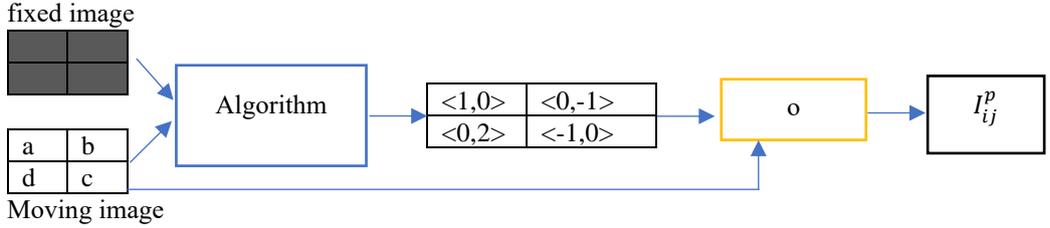

Figure 6. A displacement field estimated by an algorithm is used to register a moving image.

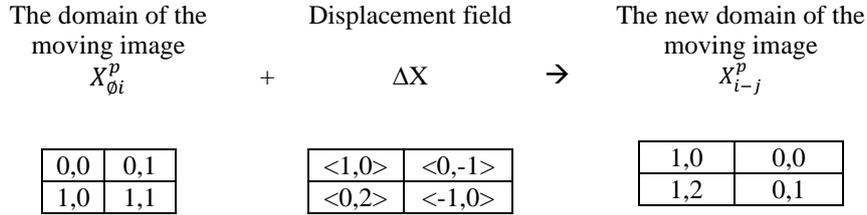

Figure 7. The displacement field transforms the domain of the moving image.

The wrapping operation 'o' shown in Figure 6 consists of a domain deformation of a moving image followed by resampling. The replacement of the domain of the moving image $X^p_{\emptyset i}$ by the domain of the wrapped image before any post-processing $X^p_{i-j}$ yields a wrapped image $I^p_{i-j}$ shown below. Figure 7 demonstrates how $X^p_{i-j}$ was obtained by the addition of ΔX to $X^p_{\emptyset i}$.

$I^p_{\emptyset i}$ = {'0,0': a, '0,1': b, '1,0': d, '1,1': c}
$I^p_{i-j}$ = {'1,0': a, '0,0': b, '1,2': d, '0,1': c}

2.3 post-processing

A domain transformation may relocate pixels to locations that violate space constraints. A constraint that is commonly violated after a domain transformation of a digital graphical image is that domain values should be uniformly distributed integers. For example, a domain value of (0.3, 0.17) violates the mentioned constraint since 0.3 and 0.17 are not integers. Such violations can be fixed in a post-processing step called 'resampling'. Resampling estimates codomain values $Y^p_{ij}$ for post-processed domain values $X^p_{ij}$ that don't violate the space constraints. The selection of $X^p_{ij}$ is known in the literature as 'grid generation'. Resampling is done under the assumption that the post-processed image <$X^p_{ij}, Y^p_{ij}$> and the preprocessed image <$X^p_{i-j}, Y^p_{i-j}$> are discrete samples from the same function/manifold. Hence $Y^p_{ij}$ can be interpolated based on $Y^p_{i-j}$. The Interpolations list includes but is not limited to linear, bilinear, and spline interpolations.

An example of post-processing is shown in Figure 8. Pixel d is out of the image boundaries. A potential fix is just to move the pixel d one step to the left to be in the empty cell at <1,1>. Another option is just to delete pixel d without filling the empty cell, another option is to fill the empty cell with the average of its surrounding pixels.

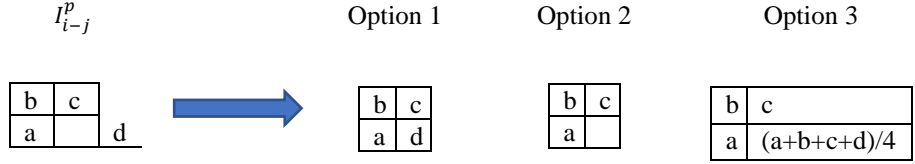

Figure 8. Examples of image interpolation

## 3.0 IR constraints

Constraints in machine learning (ML) are assumptions that limit the search space for a solution to a problem. Such assumptions are vital for generalization beyond the known examples. When a true solution is unknown, all solutions that ML models provide become equal if no assumptions are made (Mitchell, 1980). ML constraints were reviewed in Goyal, A., & Bengio, Y. (2022) and considered essential to the development of a higher level of machine intelligence – the next generation of deep learning. An example of an ML constraint is the rotation equivariance assumption which implies that a rotation of an input entails a similar rotation of the output. Geometric deep learning (GDL) was introduced in Bronstein et al., (2021) as a framework that studies the incorporation of constraints into neural network architectures based on unified geometric principles (e.g., symmetry). An example of the incorporation of priors through wavelets was done by Oyallon & Mallat (2015).

Three common IR constraints were shown to be invalid in Section 2. This section introduces IR constraints that were common among IR methods.

1- **The proximity constraint**

The proximity constraint assumes a space k in which correspondent elements are in proximity, as in Equation 3.

$$\text{Prox}(T_{ik}(I^p_{\emptyset i}), T_{jk}(I^p_{\emptyset j})) \approx 0 \qquad (3)$$

A special case of the proximity constraint is when Prox() is the distance between points in a Euclidean space. Equation 3 in this special case becomes as shown in Equation 4. The special proximity constraint assumes that, for images with a correspondence relation between them, there is a space K in which the correspondent points will be located at the same location or very close to each other.

$$T_{ik}(xe^p_{\emptyset i}, ye^p_{\emptyset i}) \approx T_{jk}(xe^p_{\emptyset j}, ye^p_{\emptyset j}) \text{ for correspondence points in } C^p_{ij} \qquad (4)$$

A limitation of the mathematical expression in Equation 4 is that 1) it measures the proximity as a distance in a correspondence space, and 2) It uses point-based proximity.

2- **Co-domain preservation constraint**

The second constraint assumes a transformation $T_{i-k}$ that transforms the domain but not the co-domain as in Equation 5.

$$I^p_{i-k} = <X^p_{i-k}, Y^p_{i-k}> = T_{i-k}(I^p_{\emptyset i}) = <T_{i-k}(X^p_{\emptyset i}), Y^p_{\emptyset i}> \qquad (5)$$

Image registration methods, to the best of the authors' knowledge so far, include a spatial transformation that affects the domain only. In other words, $T_{i-k}$ relocates pixels/voxels without changing their colors/values. In general, deep learning architectures used for MIR impose the second constraint explicitly. A deep learning model takes images as input and produces a deformation field. Then the deformation field adjusts the domain of an image using a spatial transformation operation to yield a registered version of the image. Co-domain values might be fine-tuned in a post-processing step (e.g., resampling) to fit the discretized representation of digital images in a Euclidean space, but that post-processing does not affect the co-domain preservation constraint.

Is there an IR method that violates the second constraint? A registration method that does not impose a spatial transformation step could be a potential falsifier. In theory, neural networks (NN) can approximate any function according to the universal approximation theorem. Hence, an end-to-end neural network can, in theory, register images without an imposed spatial transformation step, but no such model yet to the best of the authors' knowledge. Hence, the second constraint holds until an end-to-end neural networks model, without an explicit or implicit spatial transformation unit, is developed in practice.

Another toy falsifier is that the registration in Figure 5 can be done by applying the following codomain transformation {a: b, b: c, c: d, d: a}, which yields the same outcome shown in Figure 5C. That transformation alters the codomain and preserves the domain. IR is possible outside the co-domain preservation constraint. One may wonder what makes the constraint holds.

### 4.0 Related survey papers

Table 3 compares this work to related survey papers that appeared in the search query in section 5. Two highly cited review papers (Zitova et al., 2003; Haskins et al., 2020) were added to the table although they were published before 2021.

It could be in the interest of novice readers to read about the history of IR and its etymology (see section 1) in addition to a simple numeric example that demonstrates the basics of IR (see section 3) since no review paper was found that addresses these parts to the best of the authors' knowledge. Advanced users could be interested in the novel constraint-based analyses of IR introduced in the previous sections. Different from other survey papers shown in Table 3 which were mainly descriptive with no or just a few equations, this survey introduced a symbolic framework of the IR components (see the nomenclature) that has been used to express tens of equations.

Zitova et al. (2003) structured their paper based on the classical IR pipeline starting with feature detection, followed by feature matching, mapping function, image transformation, and resampling.

Haskins, et al. (2020) tracked the development of MIR algorithms covering 1) deep iterative methods that are based on similarity estimation, 2) supervised transformation estimation which entails ground truth labels that are not easily affordable, and 3) unsupervised transformation estimation methods which overcome the challenge of ground truth labels. Finally, 4) weakly supervised approaches were discussed.

Chen, X. et al. (2021) first provided a framework for image registration. Then explained the basic units of DL and reviewed DL methods such as deep similarity, supervised, unsupervised, weakly supervised, and RL. The authors discussed the challenges of MIR: 1) different preprocessing steps lead to different results, 2) a few studies quantify the uncertainty of predicted registration, and 3) limited data (small scale). Finally, possible research directions were highlighted: 1) hybrid models (classical methods and deep learning), and 2) Boosting MIR performance with priors.

Table 3 Comparison of related MIR survey papers.

| Paper | Review period | Taxonomy/ survey criterion | Scope limit | Etymology | Equations | symbolic framework |
|---|---|---|---|---|---|---|
| This | 2021-2022 | See the Taxonomy section and appendix | - | Y | Y | Y |
| (Zitova et al., 2003) | 1992-2002 | - feature-based, intensity-based | - | N | Y | N |
| (Haskins, et al. 2020) | 2012-2020 | - MIR algorithms: deep iterative, supervised, unsupervised,<br>- Deformability: rigid, deformable<br>- Modality type: MR, CT,...<br>- ROI<br>- Dataset: real, synthetic<br>- Loss function | Intensity-based | N | N | N |
| (Chen, X. et al., 2021) | 2013-2021 | - MIR algorithms: Deep similarity, supervised, unsupervised, weakly supervised, RL<br>- Model<br>- ROI | | N | Y | N |
| | | - Modality: unimodal, multimodal<br>- Modality type: MR, CT, …<br>- Dimensionality: 2D, 3D<br>- link to code<br>- Datasets | | | | |
| (Dossun et al., 2022) | 2010-2022 | - Evaluation metrics: Overlap (e.g., DSC), Volume (e.g., \|J\|), information theory (e.g., mutual information), probabilistic (e.g., correlation), distance based<br>- MIR tool: commercial, opensource, in-house<br>- MIR algorithms: DL, Atlas, ….<br>- Evaluation metrics<br>- Threshold<br>- Groundtruth (# observers)<br>- Dosimetric analysis (Y/N)<br>- Correlation among metrics (Y/N)<br>- Year of publication | Deformable MIR in radiotherapy treatment | N | N | N |
| (Abbasi et al., 2022) | 2013-2021 | - Deformability: rigid, deformable<br>- Modality type: MR, CT,...<br>- ROI<br>- Datasets<br>- Model<br>- Similarity metrics<br>- Evaluation metrics | Unsupervised | N | N | N |
| (Xiao et al., 2021) | 2016-2020 | - ROI<br>- Modality type: MR, CT,...<br>- Evaluation metrics.<br>- Datasets<br>- Deformability: rigid, deformable<br>- Method: deep iterative, supervised, unsupervised. | 3D | N | Y | N |
| (Chen, X., Wang et al, 2022) | 2016-2019 | -Year<br>-Application<br>-Model<br>-Dataset<br>-Contributions/ highlights | | N | N | N |
| (Huang et al., 2022) | 1997-2020 | - MIR algorithms: deep iterative, supervised, unsupervised.<br>-Tumor type<br>- Modality<br>- Model<br>- Evaluation metrics.<br>- Result in numbers<br>- Dataset size | Brain tumor | N | N | N |
| (Decuyper et al., 2021) | 2016-2020 | - MIR algo.: deep similarity metrics, supervised, unsupervised, RL<br>- Summary | Nuclear medicine and Radiology | N | Y | N |
| (Zhang, Y. et al., 2021) | No info | - Features: internal (image)/ external (info beyond image such as patient age)<br>- Ground truth: expert labeling, simulation | Breast cancer | N | N | N |

Dossun et al. (2022) reviewed the performance of deformable IR in radiotherapy treatments in real patients. First, the scope of the paper and the paper selection process were explained. Then a taxonomy of MIR evaluation metrics was mentioned but no explanation or formula was provided. A table of 7 pages compared the surveyed papers. Then, statistics and figures summarized the results showing, for example, that the distribution of the ROIs was 36% for the prostate (Bashkanov et al., 2021; Yang, Q. et al., 2021; Yang et al., 2022), 33% for the head and neck, and 26% for the thorax. Another figure showed the most frequent evaluation metrics ordered as the following: DSC > HD > TRE.

Abbasi et al. (2022) reviewed the evaluation metrics of unsupervised MIR in a sample of 55 papers. The statistics showed that: 1) the majority of papers were handling unimodal registration (82%), 2) a private dataset was more likely to be used than a publicly available dataset, 3) most papers worked with MR images (61%), and 4) the most researched ROI was the brain at 44%, then the heart at 15%.

Xiao et al. (2021) started with a brief introduction to deep learning, then provided a statistical analysis of a selected sample of 3D MIR papers that covered the distribution of ROI (Brain 40%, lung: 24%), modality (MR-MR: 46%, CT-CT: 24%), MIR methods: (unsupervised: 43%, supervised: 40%, deep iterative: 19%), and evaluation metrics (74% label based, 18% deformation based, 12% image-based). The MIR methods were reviewed based on the taxonomy (deep iterative methods, supervised, and unsupervised).

Chen, X., Wang, et al. (2022) reviewed medical image analysis covering four areas: image classification, detection, segmentation, and registration. First, the paper gave an overview of deep learning and its methods:

supervised, unsupervised, and semi-supervised. Then it addressed ideas of DL that were shown to improve the outcomes: attention, involvement of domain knowledge, and uncertainty estimation. Then the paper briefly reviewed classification, detection, segmentation, and registration. Finally, the paper highlighted ideas for future improvement that included the idea of a fully end-to-end deep learning model for MIR. In addition to the incorporation of domain knowledge. They also highlighted important points for large-scale applications of deep learning in clinical settings such as having large datasets publicly available as well as producible codes. They also highlighted the need for more clinical-based evaluation and the involvement of domain experts from the medical field in the evaluation rather than limiting the evaluation to theoretical evaluation metrics.

Huang et al. (2022) reviewed AI applications in brain tumor imaging from a medical practitioner's perspective. They pointed out the lack and the need for studies about the use of AI tools in routine clinical practice to characterize the validity and utility of the developed AI tools.

Zhang, Y. et al. (2021) elaborated on AI registration success, and highlighted challenges 1) the lack of large databases with precise annotation, 2) the need for guidance from medical experts in some cases, 3) having different opinions of experts in the case of some ambiguous images. 4) excluding non-imaging data of the patient, like age, and medical history, and 5) the interpretability of AI models.

Decuyper et al. (2021) started with an explanation of DL components covering neural network layers (CNNs, activations, normalization, pooling, and dropout), and DL architecture (e.g., Resnet, GANs, U-Net). Then the paper explained medical image acquisition and reconstruction. After a brief elaboration on IR categories, the paper elaborated on their challenges: 1) traditional iterative methods work well with unimodal images but poorly with multimodal images, or in the presence of noise, 2) deep iterative methods imply non-convex optimization that is difficult to converge, 3) In RL, deformable transformation results in a high dimensional space of possible actions, which makes it computationally difficult to train RL agents. Most previous works dealt with rigid transformation (low dimensional search space), 4) supervised learning approaches need ground-truth labels, and 5) unsupervised approaches face difficulty in back-propagating the gradients due to the multiple different steps. Finally, specific application areas were reviewed: chest pathology, breast cancer, cardiovascular diseases, abdominal diseases, neurological diseases, and whole-body imaging.

This work reviewed MIR. The list of surveyed work was collected by searching using the keywords medical image registration in the Scopus database. The search query was limited to open-access MIR papers written in English and published between 2021 and 2022. The number of retrieved records was 270, out of which 38 papers were excluded based on the abstract for their irrelevance (e.g., they are about medical images but not MIR). Other 41 papers were excluded as the authors could not find open-access versions of those papers as of December 2022. Out of 191 papers, 96 have been reviewed at the time of writing this draft in addition to 10 papers published before 2021. The research questions and sub-questions of this work are shown in Table 4. The outcomes of the survey are summarized in the appendix.

> **Scopus search query:**
>
> Key words in title/abstract/keywords: medical "image registration"
> Source type: Journal or conference proceedings
> Year: 2021 and 2022
> Language: English
> open access

Table 4. The questions of interest in this survey

| Research question | Sub-questions |
|---|---|
| What was the research pipeline of MIR in 2021 and 2022? | What were the proposed approaches of MIR? |
| | What were the evaluation criteria? |
| | What MIR datasets were used |
| | What were the applications/use cases of MIR? |

### 5.0 Taxonomies

A registration algorithm consists of a set of assumptions (prior knowledge), and a margin of uncertainty (the unknown part), which is expressed using variables (e.g., model parameters). For example, if a programmer knows exactly how to register any images in a similar way to having a formula that finds the roots of any quadratic equation, then s/he will just embed that prior knowledge (the formulae) in the code. However, there is no such a generic formula yet for most IR cases. Accordingly, variables are made and adjusted using an optimization method.

5.1 Deformation types

Transformation functions in MIR can be categorized based on their deformability into rigid, affine, and deformable transformations as shown in Figure 9.

In physics, the shape and size of a rigid body do not change under force. When you push a small solid steel bar, the location and/or the orientation of the bar may change, but the bar itself remains the same (e.g., the same mass, shape, and size). Likewise, a rigid transformation preserves the distances between every pair of points. Accordingly, rotations and translations are rigid transformations or proper rigid transformations in the distinction of reflections which are called improper rigid transformations as they do not preserve the handedness.

A rigid transformation $T_{ij}$ preserves the distances between any two points on the object of interest, such that the constraint $\left\|xk_{\emptyset i}^p - xl_{\emptyset i}^p\right\| = \left\|xk_{ij}^p - xl_{ij}^p\right\|$ holds for every pair of points k, l ∈ the set $M_p$. A rigid transformation can be expressed as in Equation 6.

$$\tilde{v} = T_{ij}(v) = \mathbf{A}\, v + b \qquad (6)$$

Where $\tilde{v}$ is a newly transformed vector after the application of a rigid transformation to a vector v, which could be a position of a point in Euclidian space. b is a translation vector, and A is an orthogonal transformation (see the appendix for definition) such as orientation.

A rigid transformation is a subcategory of a bigger group of transformations called affine transformations. Affine transformations preserve parallelism and lines, but no constraints on the preservation of distances. Thus, it can be expressed as in Equation 6 above used earlier for rigid transformation except that **A** is a linear transformation/matrix with no orthogonality constraint. In an affine registration, the transformation $T_{ij}$ imposes the constraint $T_{ij}\left(xk_{\emptyset i}^p - xl_{\emptyset i}^p\right) = T_{ij}\left(xk_{\emptyset i}^p\right) - T_{ij}\left(xl_{\emptyset i}^p\right) = xk_{ij}^p - xl_{ij}^p$ for every point k, l ∈ the set $M_p$. Scaling and shear mapping are examples of an affine, but not rigid, transformation.

The formula of a 2D proper rigid transformation (rotation and translation) is shown in Equation 7. The variables are the rotation angle θ, the translation on the x-axis $bx$, and the translation on the y-axis $by$.

$$\tilde{v} = \begin{bmatrix}\cos(\theta) & -\sin(\theta) \\ \sin(\theta) & \cos(\theta)\end{bmatrix} v + \begin{bmatrix} bx \\ by \end{bmatrix} \qquad (7)$$

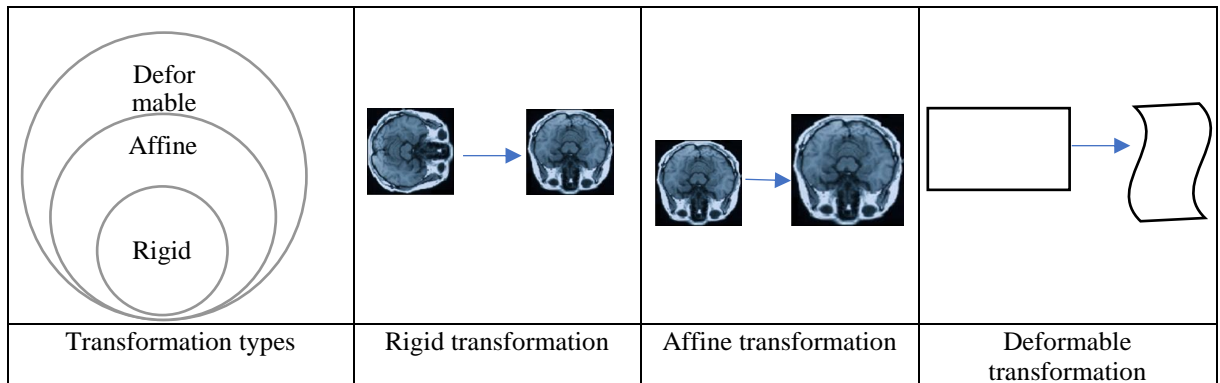

Figure 9. Examples of deformation types

The formula of a proper rigid transformation in a 3D space consists of 6 unknown variables: 3 rotation angles (θx, θy, θz), and 3 translations (bx, by, bz) as shown in Equation 8, where the subscriptions x, y, z are 3 perpendicular coordinates.

Deep learning in medical image registration: introduction and survey

$$\tilde{v} = \begin{bmatrix} 1 & 0 & 0 \\ 0 & \cos(\theta x) & -\sin(\theta x) \\ 0 & \sin(\theta x) & \cos(\theta x) \end{bmatrix} \begin{bmatrix} \cos(\theta y) & 0 & \sin(\theta y) \\ 0 & 1 & 0 \\ -\sin(\theta y) & 0 & \cos(\theta y) \end{bmatrix} \begin{bmatrix} \cos(\theta z) & -\sin(\theta z) & 0 \\ \sin(\theta z) & \cos(\theta z) & 0 \\ 0 & 0 & 1 \end{bmatrix} v + \begin{bmatrix} bx \\ by \\ bz \end{bmatrix} \quad (8)$$

Transformations that do not preserve the rigidity or affinity constraints are called deformable transformations.

5.2 Optimization phase

Image registration entails an optimization step in which a model's parameters are adjusted to minimize/maximize an objective function. Optimization can occur, as shown in Figure 10, 1) during the development phase as in DL approaches, or 2) during the running phase such as in iterative methods, or 3) in both e.g., active learning approaches, or a test-time training as called in (Zhu et al., 2021). The objective function of MIR is expressed in Equation 9 as a weighted sum of two components: the first quantifies the registration error that represents the proximity between the predicted registration and the correct one, and the second is a regularization component.

$$\text{Loss} = \text{registration\_error} + \text{regularization} \quad (9)$$

The optimization methods such as gradient descent, evolutionary algorithms, and search are iterative. Hence the optimization step adds a time overhead to the phase in which it takes place. Thus, DL approaches take a long training time, but shorter registration time.

Approaches that run optimization in both phases aim at further improving the registration despite a slight increase in the computation time. To reduce the run-time overhead, the bulk optimization of the model parameters occurs in the training phase while just slight finetuning occurs during the run phase to customize the results (Zhu et al., 2021).

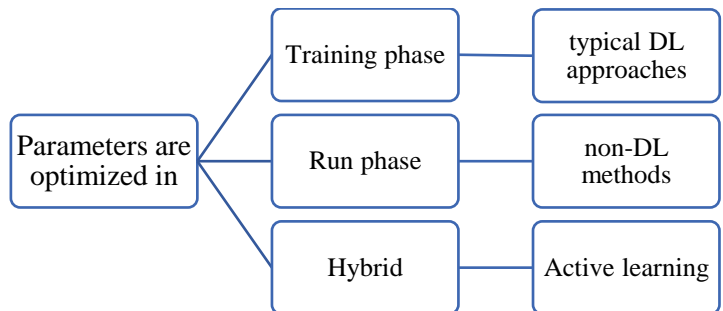

Figure 10. Optimization phase

5.3 MIR algorithms

This section discusses selected registration algorithms. Mainly the algorithms that were used as baselines against which the performance of a new algorithm is compared. A taxonomy of MIR algorithms is shown in Figure 11.

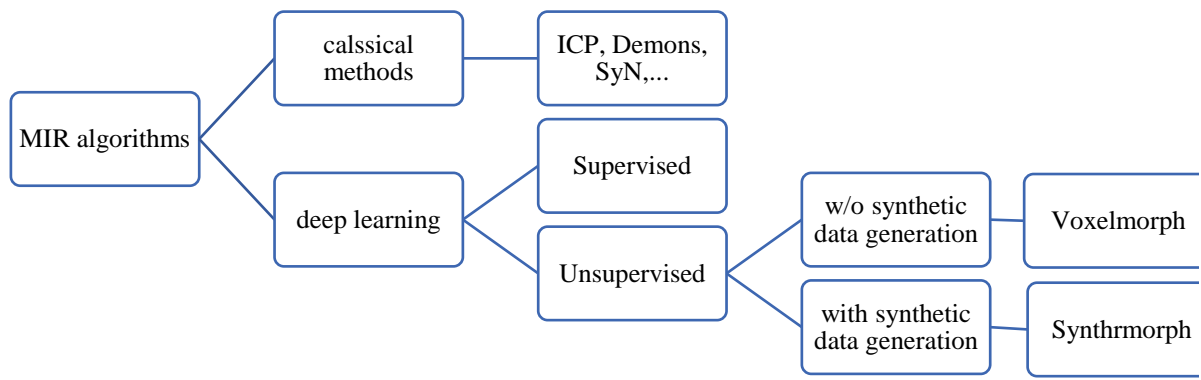

Figure 11. MIR methods taxonomy

*5.3.1 Deep learning approaches*
Deep learning approaches use multiple layers of neural networks. Neural networks can estimate the transformation function in the registration problem entirely using unknown variables (called neurons). Hence, the transformation function in this case is considered an implicit function in distinction with explicit transformation functions which assumes a tractable formula of the transformation functions such as rigid transformations shown in Equations 6-8. DL approaches were also called earlier "non-parametric methods".

a. directly supervised deep learning approaches.

The diagram of directly supervised image registration approaches is shown in Figure 12. Initially, input images are fed to neural networks which produce a registration field. The registration field is applied to the fixed image to relocate its pixels in a process called spatial transformation represented as a yellow circle in the figures below. An example of a supervised registration can be seen in (Lee et al., 2022).

The main question is how neural networks learn to estimate the registration field. In the directly supervised approach, A ground truth label is provided during the training phase. The ground truth label could be the registration field as shown in Figure 12 (left), or the wrapped image as shown in Figure 12 (right). A challenge of directly supervised MIR approaches is their need for ground truth labels, which entails medical experts annotating a large number of images. To overcome ground truth labels, unsupervised MIR has been proposed.

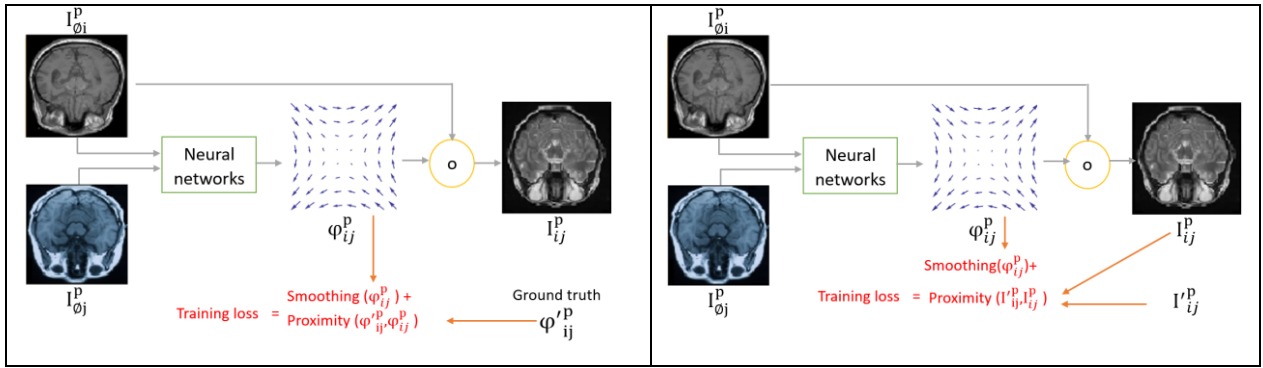

Figure 12. Supervised MIR approaches: supervision using ground truth output image (right), and supervision using ground truth registration field (left)

b. Unsupervised deep learning approach: Voxelmorph

Unsupervised MIR approaches do not entail an external supervision signal. Instead, the fixed image (input) was assumed to replace the ground truth label of the registered image $< X_{ij}^{p'}, Y_{ij}^{p'} > \approx < X_{\emptyset j}^{p}, Y_{\emptyset j}^{p} >$ as in Voxelmorph (Balakrishnan et el., 2019). This assumption is useful when the fixed image and the moving image have similar modalities/co-domains. However, the assumption may not work well if the fixed image and the registered image are of different modalities (e.g., one is 3D MRI, and the other is 2D X-ray) unless a way is developed to bridge the gap between the two modalities. This has been reported by the results shown in Synthmorph (Hoffmann et al., 2021). Even for images of the same modality, co-domain dissimilarities can be a problem with this approach. For example, if the contrast of the fixed image is different than that of the moving image, then the mean square error $MSE(Y_{\emptyset j}^{p}, Y_{ij}^{p})$ may not represent the error adequately. However, another loss function like cross-correlation "CC" is more resilient against the contrast problem than MSE due to its scale invariance property. CC (Y1, Y2) = CC (Y1, α×Y2) where α is a scale number.

MIR using Voxelmorph yielded results much faster than non-deep learning MIR methods without degradation of the registration quality. Voxelmorph cut the registration runtime to minutes/seconds compared to hours needed by non-deep learning methods used before Voxelmorph. Voxelmorph superseded non-deep learning methods when segmentation labels were added to the registration.

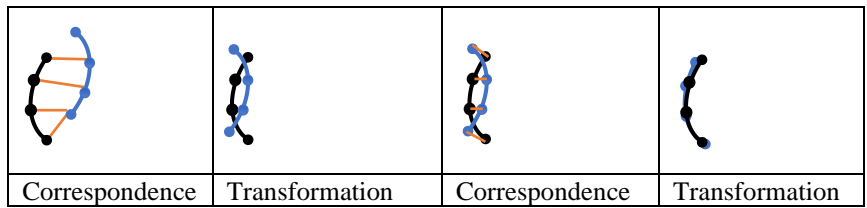

Figure 14. A demonstration of ICP registration.

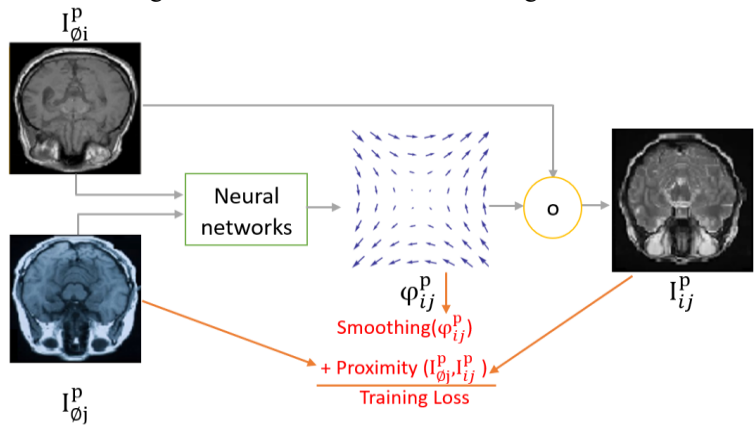

Figure 13. Unsupervised MIR approach

c. Unsupervised approach with synthetic data: Synthrmorph

If it is difficult to get ground-truth labels, why not generate them? Synthmorph (Hoffmann et al., 2021) proposed training Voxelmorph on synthetic data (randomly generated fixed and moving images). Synthmorph generated images in two steps, first segmentation labels were generated randomly, then fixed, and moving images were generated given the segmentation label. The results yielded by Synthrmorph were superior to classical methods even when the images were of different modalities.

*5.3.2 non-deep learning methods:*

MIR methods that do not involve deep neural networks are called 'non-deep learning methods', 'classical methods', or 'iterative methods.'

a. Iterative Closest Point (ICP)

ICP (Arun, 1987; Estépar, 2004; Bouaziz, 2013) alternates between two goals: the establishment of a correspondence $C_{ij}^{pq}$, and finding a transformation $T_{ij}$ that optimizes a loss function. A loss function quantifies the quality of a registration (see section 8). A demonstration of the ICP process is shown in Figure 13. Let the moving image be a blue line of 4 marked points, and the fixed image a similar black line. The loss function can be a point-wise Euclidean distance. First, 1) a correspondence is established between the points on each line such that each point is matched with its closest neighboring point. Notice that the correspondence is not 1-to-1 as the two bottom black points are matched with the same point, and the top blue point is not matched, 2) the blue line was translated to minimize the distance between the two lines, 3) another correspondence was found (1-to-1 correspondence this time), and 4) the black line was transformed (rotation and translation) based on the new correspondence.

ICP, like other iterative approaches, takes longer registration time than DL approaches. The establishment of a correspondence between nearest neighbors is straightforward but not always optimal and it sticks in local optima.

b. Demons
A deformable IR approach was proposed by Thirion (1996). The name of the Demons approach was influenced by Maxwell's Demons paradox in Thermodynamics. Maxwell assumed a membrane that allows particles of type A to pass in one direction, while particles of type B can pass in the opposite direction, which will end up having all particles of type A on one side of the membrane and particles of type B on the other side as shown in Figure 15. That state of organized particles corresponds to a decrease in entropy, which contradicts the second law of thermodynamics. The solution to that paradox was that the demons generate entropy to organize the particles resulting in a greater total entropy than that was before the separation of the particles.

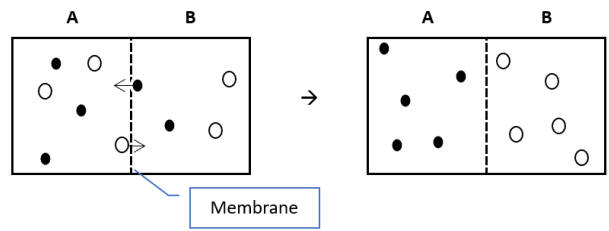

Figure 15. Maxwell's membrane with demons

Influenced by Maxwell's demons, Thirion suggested distributing particles (demons) on the boundaries of an object (see Figure 16) such that a demon will push locally either inside or outside the object based on a prediction of a binary classifier. It has been shown that what Thirion's demons do is object matching using optical flow.

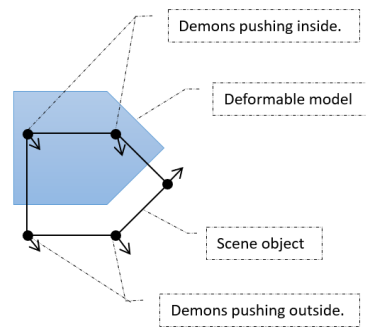

Figure 16. How demons work as explained in Thirion, J. P. (1996, June)

c. Symmetric Image normalization (SyN)
The main idea of SyN is to assume a symmetric and invertible transformation. Instead of transforming space i to j, SyN symmetrically transforms both space i & space j to an intermediate space such that $T_{jk} = T_{ik}^{-1}$. In this case $T_{ik}$ can be seen as half a step forward towards space j, and $T_{jk}$ is half a step backward towards i (see Figure 17). The symmetric invertibility constraint of SyN can be expressed as in Equation 10

$$\exists k \in space: \ T_{ij}(I_i) = T_{kj}((T_{ik}(I_i)), where \ T_{kj} = T_{ik}^{-1} \qquad (10)$$

SyN was shown to supersede Demons in providing correlated results with human experts (Avants et al., 2008).

# Deep learning in medical image registration: introduction and survey

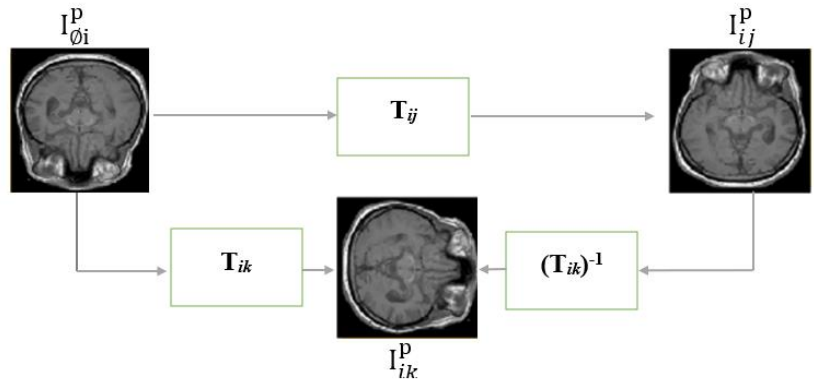

Figure 17. A demonstration of symmetry in SyN. A Transformation Tij, which rotates an image 180 degrees counterclockwise, can be decomposed into 2 symmetric rotations each of 90 degrees.

d. Registration software tools.

NiftyReg is a publicly available software for image registration. The software was developed initially by University College London and then King's College London. The software uses two methods: 1) Reg Aladin, which is a block matching algorithm for global registration based on Ourselin et. al. (2001). 2) RegF3D (fast free form deformation) based on Modat et al (2010).

Advance Normalization Tools (ANT) is another stable software for MIR and statistical analysis. ANT yields stable results such that the registration does not change every time the software is run (Avants et al., 2014). A Python version of NiftyReg and ANTs was wrapped in a package called Nipype (Neuroimaging in Python pipelines & interfaces).

ANTs on Github: https://github.com/ANTsX/ANTs

Chen, T. et al., (2002) compared three registration tools: SPM12, FSL, and AFNI. *SPM12* was recommended for novice users in the area of medical image analysis. It provided stable outcome images of "maximum contrast information" needed for tumor diagnosis. *AFNI* was recommended for advanced users and researchers due to the advanced capabilities needed for tasks such as volume estimation. *FSL* was considered for mid-level users.

5.4 Correspondence space

MIR alignment occurs in a correspondence space k. The correspondence space can be the space in which an input image is located (internal), or it can be a new space (external). MIR in an internal correspondence space has been the most common among MIR methods. Examples of MIR in an internal space can be seen in the methods mentioned earlier, which included a transformation from the space of a moving image (i) to the space of the fixed image (j). An example of MIR in external space is Atlas-based registration (Wang, Z. et al., 2022).

**Atlas-based registration**

An Atlas is a standard or a reference image that represents a population of images. One way to form an Atlas of a brain is by finding the average image of a population of brain images, which is expected to be smooth and symmetrical. However, that is not the only way. (Dey et al., 2021) suggested an atlas generated by GANS. Another way to form an atlas is by IR in an external correspondence space. An example of atlas-based registration is the Aladdin framework (Ding, Z. et al., 2022) shown in Figure 18. Aladdin transformations are bidirectional and invertible.

Deep learning in medical image registration: introduction and survey

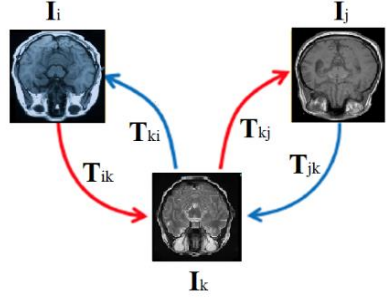

Figure 18. Invertible bidirectional Atlas-based transformations

- Invertibility: for a transformation $T_{ij}$, there is an inverse transformation $T_{ij}^{-1}$
- Bidirectionality: A bidirectional registration maps spaces in both directions from i to k and vice versa ($T_{i\leftrightarrow k}: T_{ik}$, and $T_{ki}$). Accordingly, a bidirectional IR model (Ding, W. et al.,2022; Andreadis et al., 2022; Ye et al.,2021) can yield two wrapped images $I_{ij}, I_{ji}$. On the other side, a unidirectional registration maps a single space i into another j but not vice versa. An example of an invertible bidirectional MIR model in an internal correspondence space, namely Inversenet (Nazib et al., 2022), is shown in Figure 19.

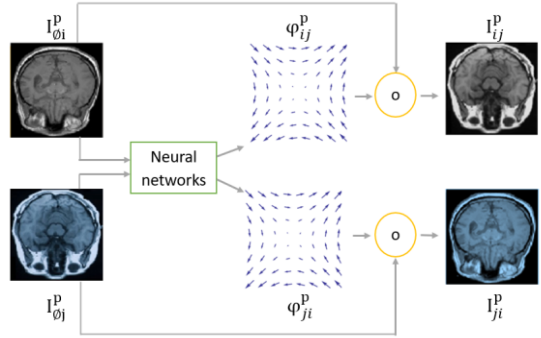

Figure 19. An example of a MIR model that estimates a transformation field and its inverse (InverseNet)

The bi-directionality in an external correspondence space enables more transformation paths between spaces given three anchors as shown in Figure 18: fixed image $I_i$, moving image $I_j$, and an external correspondence space/Atlas $I_k$. Potential transformation paths were expressed in Equations 11-17 below. The dissimilarities between the left and right sides of the equations below were used as losses of an MIR model (Ding, Z. et al. 2022).

$$T_{ki}(T_{ik}(I_{\emptyset i})) = I_{\emptyset i} \qquad (11)$$
$$T_{ki}(T_{jk}(I_{\emptyset j})) \approx I_{\emptyset i} \qquad (12)$$
$$T_{ki}(I_{\emptyset j}) = I_{ki} \approx I_{\emptyset i} \qquad (13)$$

$$T_{kj}(T_{jk}(I_{\emptyset j})) = I_{\emptyset j} \qquad (14)$$
$$T_{kj}(T_{ik}(I_{\emptyset i})) \approx I_{\emptyset j} \qquad (15)$$
$$T_{kj}(I_{\emptyset k}) = I_{kj} \approx I_{\emptyset j} \qquad (16)$$

$$T_{ki}(I_{\emptyset i}) = T_{jk}(I_{\emptyset j}) \qquad (17)$$

Figure 20 illustrates diagrams of IR in an internal correspondence space (left) and an external correspondence space (right).

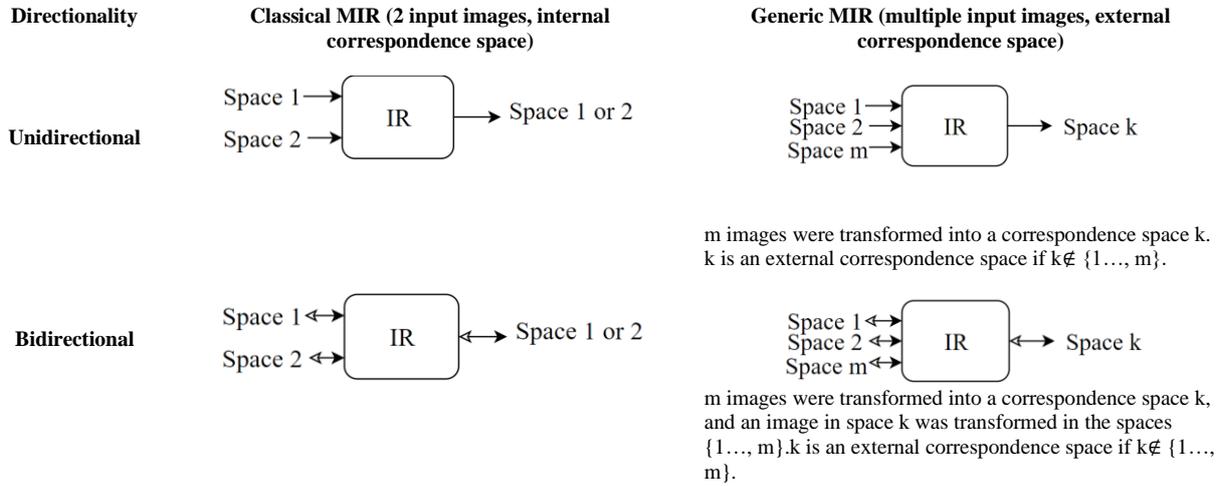

Figure 20. diagrams of unidirectional and bidirectional IR in internal and external correspondence spaces

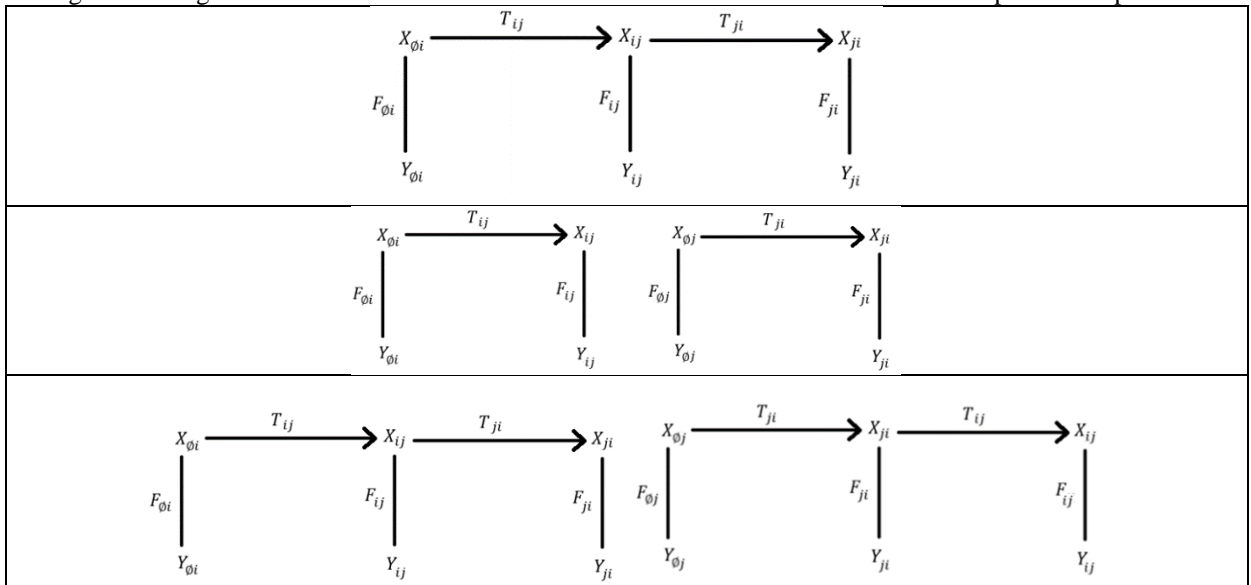

Figure 21. Space diagrams of bidirectional IR models. A transformation T in the space diagram can be neural networks followed by a spatial transformation. The parameter on the left side of a transformation in the space diagrams represents the domain of the image to be transformed into another space, not the inputs of the neural networks.

5.5 Correspondence relation

Correspondence relations can be categorized into isomorphic and non-isomorphic. Isomorphism entails a one-to-one correspondence relation between images. A special case of isomorphism is diffeomorphism which entails invertible and differentiable transformation. An example of non-isomorphism is a change of the topology such as that shown in Figure 22 B. A special case of non-isomorphism is a many-to-many correspondence as in metamorphism.

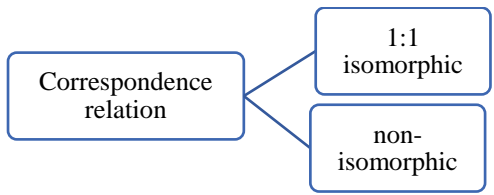

Figure 22 A. Correspondence relation taxonomy

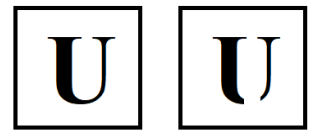

Figure 22 B. An example of metamorphism

The spatial transformation unit imposes isomorphism, since the registration field just maps a single pixel from one location to another single point only, which is a 1:1 correspondence. However, the resampling step can affect the 1:1 correspondence relation, for example, if two nearby points are merged in the target image, which makes metamorphism possible but no guarantees. Diffeomorphism can be achieved by an integral $\int$ before a spatial transformation.

Metamorphosis (Maillard et al., 2022) is a deep learning model that addresses metamorphic registration. Metamorphosis estimated the wrapped image without an explicit spatial transformation unit. However, alternative constraints were added as 2 equations embedded in the network as layers. However, no information if a spatial transformation holds implicitly. Metamorphosis superseded diffeomorphic registration methods especially when the ground truth correspondence was metamorphic. However, its runtime was 10-20 times that of Voxelmorph. The runtime is defined in section 8 (evaluation measures).

5.6 Multistage image registration

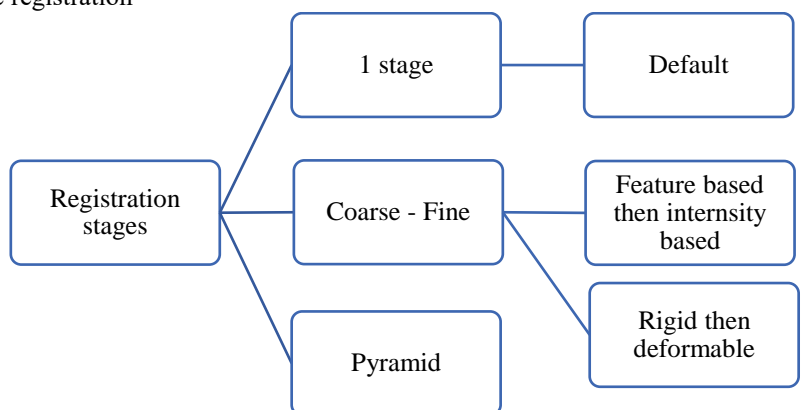

Figure 23. Taxonomy of image registration stages

Instead of solving the registration problem for high-resolution images entirely in a big dimensional space, the registration problem can be conquered into multiple registration problems of various scales. Figure 23 shows a taxonomy of multistage image registration. Multistage MIR approaches save computational resources and time in addition to the enhancement of registration results.

*7.6.1 Coarse-fine registration:*
A coarse-fine registration (Himthani et al., 2022; Naik et al., 2022; Saadat et al., 2022; Van Houtte et al.,2022) consists of two stages: The first stage is called coarse registration, which aims at finding a fast registration solution but not optimal. That solution is fine-tuned later in the second stage. For example, the coarse registration could be an affine registration that aligns the position and orientation while the fine-tuned registration could be a deformable registration method that aligns deformed parts.

The parameters of a rigid transformation of a high-resolution image can be found using a downscaled version of the image, which would save computation time and energy. The parameters of a rigid transformation are either independent of the scale (e.g., rotation) or linearly dependent (translations). Assume an image of 1000x1000 pixels and its lower resolution version of 100x100 (downscaling by 10). Scaling does not affect angles, hence if an object is rotated by 30 degrees in the downscaled image, it will be also rotated by the same angle in the high-resolution image. However, distances between objects do change according to a fixed scale. If the distance between 2 objects in the low-resolution image is 25 units, then the equivalent distance in the high-resolution image will be 10×25 = 250, where 10 is the scaling ratio between the two images. Hence a solution for a rigid registration problem can be solved in a downscaled version of the images and then transferred to the higher resolution image.

*7.6.2 Pyramid image registration.*
A pyramid consists of multi-scale images, where registration occurs at multiple stages. The idea of a pyramid representation has been well-studied in classical computer vision (Adelson et al., 1984) and utilized later in deep learning architectures such as Pyramid GANs (Denton et al., 2015; Lai et al., 2017). A pyramid registration (Wang et al., 2022; Chen, J. et al. 2022, Zhang, L. et al., 2021; Zhang, G. et al., 2021) starts with a downscaled version of the moving image followed by several operations of registration and upscaling as shown in Figure 24. After every registration step, the proximity between the wrapped image and the downscaled fixed image improves. Multi-stage registration can be seen as a sort of curriculum learning (Bengio et al., 2009; Burduja et al., 2021) such that the first stages learn to solve easier problems and later stages learn the more difficult tasks. In (Wang, C. et al., 2022), both pyramid and coarse-fine registration were used.

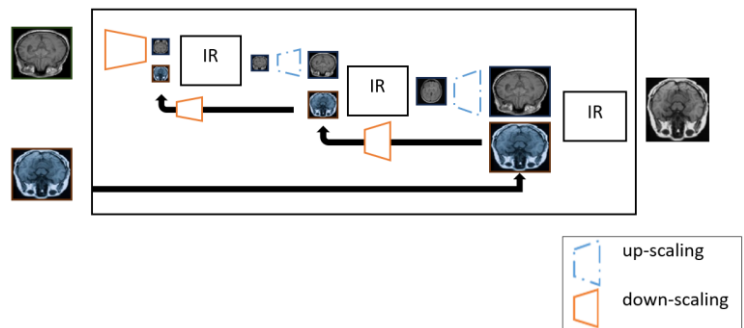

Figure 24. Pyramid registration of three stages

5.7 Space Geometry

A taxonomy of spaces has been proposed GDL is shown in Figure 25. A space can be Euclidean-like RGB images (pixels distributed regularly in a rectangle). Non-Euclidean spaces are represented in sets, graphs, meshes, or manifolds. Examples of MIR for non-Euclidean data, specifically 3D point clouds, have been presented in (Terpstra et al., 2022; Su et al., 2021).

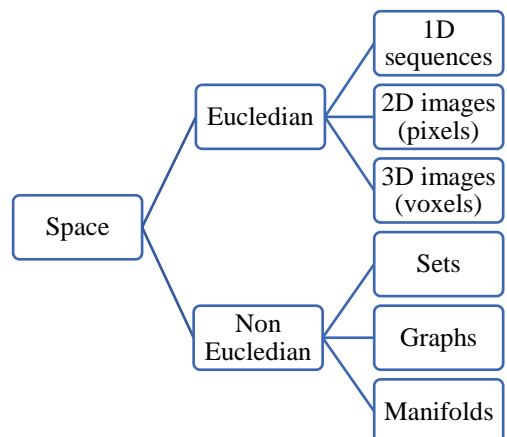

Figure 25. Space geometry taxonomy

5.8 Other taxonomies

*5.8.1 Feature-based and pixel-based*

Feature-based and pixel-based taxonomy depends on the type of inputs to the registration algorithm. A feature-based registration (Saiti et al., 2022; Santarossa et al., 2022; Wang, H. et al., 2022; Liu et al., 2021) involves an explicit feature extraction or selection, thus the input to the registration algorithm is not the image itself but representative features of that image such as its histogram (Ban et al., 2022). In pixel-based approaches, images are fed directly to the model without feature extraction. In general, DL registration approaches are pixel-based as neural networks can extract features implicitly. Some works included both feature and pixels (Ringel et al., 2022; Yang, Y. et al., 2021).

*5.8.2 Medical imaging modalities*

Medical imaging modalities are imaging techniques used to visualize the body and its components (Kasban et al., 2015). The main medical imaging modalities in MIR are:

    a.   X-ray

X-ray uses ionizing radiation (X-rays) to produce two-dimensional images of bones and dense tissues. X-rays are absorbed differently by different tissues, allowing visualization of structures like bones, lungs, and some organs.

Deep learning in medical image registration: introduction and survey

X-rays are quick and relatively inexpensive, thus suitable for some diagnostic purposes, such as detecting fractures, lung infections, and dental issues. However, they provide limited details about soft tissues.

b. Computed Tomography (CT) scan

CT scan, also known as CAT (Computerized Axial Tomography), is a non-invasive imaging technique that uses X-rays to create detailed cross-sectional images of the body. A CT scan provides a more detailed view of bones, blood vessels, and solid organs compared to traditional X-rays. It is especially useful for imaging areas like the brain, chest, abdomen, and pelvis. However, they involve exposure to ionizing radiation, and repeated scans should be minimized to reduce radiation exposure. During a CT scan, the X-ray source rotates around the patient, and multiple X-ray images are captured from different angles. These images are then processed by a computer to create cross-sectional slices, allowing doctors to visualize the body in detail. CT scans are commonly used in emergencies, trauma cases, and cancer staging, among other applications. MIR of CT images was reported in (Dida et al., 2022; Gao et al., 2022).

c. Magnetic Resonance Imaging (MR)

MRI uses strong magnetic fields and radio waves to create detailed images of tissues, organs, and the central nervous system. It provides high-resolution, multi-planar images, making it ideal for diagnosing conditions in the brain, spinal cord, muscles, and joints. MRI does not use ionizing radiation, which makes it safer, but it can be more time-consuming and expensive compared to X-rays and CT scans. MIR of MR images was reported in (Li et al., 2022; Meng et al., 2022; Himthani et al., 2022; Kujur et al., 2022; Wu et al., 2022; Ashfaq et al., 2022).

d. Ultrasound (US)

Ultrasound, also known as sonography, uses high-frequency sound waves to create real-time images of internal organs and structures. It is commonly used for imaging the abdomen, pelvis, heart, and developing fetus during pregnancy. Ultrasound is non-invasive and does not involve ionizing radiation. It provides real-time imaging and is excellent for assessing blood flow and certain soft tissue abnormalities. However, it may not provide as detailed images as MRI and CT.

e. Positron Emission Tomography (PET)

PET is a functional imaging technique that provides information about metabolic activity and cellular function. It involves the injection of a radioactive tracer that emits positrons. The interaction between the tracer and tissues produces gamma rays, which are detected by the PET scanner. PET is valuable in oncology (cancer imaging) and neurology (e.g., detecting Alzheimer's disease). PET can be combined with CT imaging to provide both functional and anatomical information in a single scan.

MIR is considered "unimodal" when there are no modality differences between the images involved in the registration process, otherwise, the registration is considered "multimodal". See Figure 26. An example of a unimodal registration is when both moving and fixed images are X-rays. An example of multimodal registration is when a fixed image is of the T1-weighted MRI modality and the moving image of the T2-weighted MRI. T2-weighted MRI enhances the signal of the water and suppresses the signal of the fatty tissue while MRI/T1 does the opposite. Examples of multimodal registration can be seen in (Van et al., 2022; Begum et al., 2022; Xu et al., 2021).

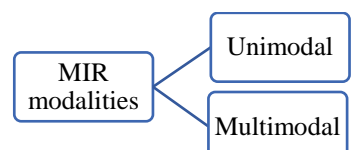

Figure 26. MIR taxonomy based on the modalities

## 6.0 Evaluation measures

IR evaluation measures can be categorized as shown in Figure 27 into 1) time-based measures that focus on the time needed to finish a task, 2) size-based measures that focus on the memory resources that an MIR algorithm occupies, 3) smoothness measures that focus on the smoothness of the registration field (expressed by Jacobian),

# Deep learning in medical image registration: introduction and survey

and 4) proximity-based measures that find the deviation of a registration outcome from the ground-truth. proximity can be expressed using distances between objects in a space, overlap between sets, or correlations between variables.

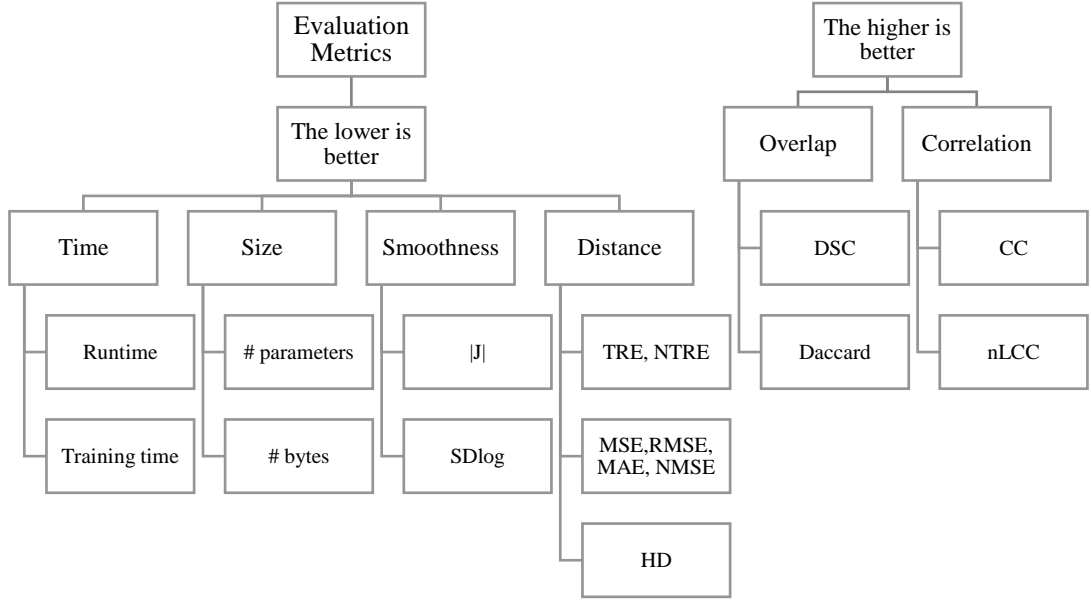

Figure 27. Evaluation Metrics

## 6.1 Time

a)  Average registration runtime RT:

The runtime (RT) is the average registration time per image. The registration time is measured from the moment $t^p_{r,start}$ at which an image p is loaded until obtaining the registered image at time $t^p_{r,end}$ including the post-processing time. See Equation 18. Where N is the number of examples in a dataset.

$$RT = \frac{\sum_{p=1}^{N}(t^p_{r,end} - t^p_{r,start})}{N} \tag{18}$$

In practice, getting the registration outcome in a short time is a desired property. The Voxelmorph algorithm, which uses deep learning for medical image registration, has shown an RT reduction from hours to seconds while keeping almost the same performance. The computation time of a registration process depends on the software as well as the hardware (Alcaín et al., 2021). Thus, a fair comparison of registration algorithms entails testing the computation time on the same hardware. The shorter RT of Voxelmorph compared to iterative approaches can be attributed partially to the hardware, where matrix multiplication processes used in DL are faster when run with a GPU. However, even on CPUs, Voxelmorph remains faster than iterative methods on a scale of minutes for voxelmorph to hours for iterative methods. The main reason for the longer RT in iterative approaches is the optimization done during the runtime, however, Voxelmorph-like approaches do not optimize the variables during the run phase, instead, all the variables are optimized in the training phase before the run time.

b)  Average training time DT is the training time divided by the number of examples in the training dataset.

## 6.2 Distance-based measures
The distance can be chosen to be between co-domain values or domain values. The distance can be measured between selected points (landmarks) or all points.

a)  Codomain distance: MSE, RMSE

The Euclidean point-wise distance between codomain values can be calculated using the mean square error (MSE), and root mean square error (RMSE) measure as in Equations 19, and 20 respectively.

$$MSE = \frac{1}{N} \sum_{p=1}^{N} \frac{1}{Lp} \sum_{e=1}^{Lp} Dist(ye'^{p}_{ij}, ye^{p}_{ij})^2 \qquad (19)$$

$$RMSE = \sqrt{MSE} \qquad (20)$$

b) Domain distance: TRE

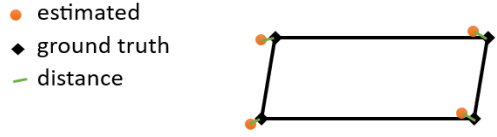

Figure 28. TRE components

TRE (target registration error) is a distance-based metric that measures the deviation between points of two domains. See the deviation between estimated and ground truth points in Figure 28. The distance is used to represent the registration error since a perfect registration would locate correspondent points ideally at the same position. In the case of ground truth labels $X'_{ij}$ and predictions $X_{ij}$, TRE is shown in Equation 21.

$$\text{TRE} = \text{RMSE}(X'_{ij}, X_{ij}) = \sqrt{\frac{1}{N} \sum_{p=1}^{N} error(M'^{p}_{ij}, M^{p}_{ij})^2} = \sqrt{\frac{1}{N} \sum_{p=1}^{N} \sum_{e=1}^{Mp} Dist(xe'^{p}_{ij}, xe^{p}_{ij})^2} \qquad (21)$$

where $xe^{p}_{ij} \in M^{p}_{ij}$, $xe'^{p}_{ij} \in M'^{p}_{ij}$, with a corrosondence $(xe'^{p}_{ij} \leftrightarrow xe^{p}_{\emptyset j})$

c) Domain distance: NTRE

TRE is affected by the scale of an image as well as the number of landmarks, the more landmarks in an image the higher the accumulative error could be. The normalized target to registration error (NTRE) is scale independent as shown in Equation 22.

$$NTRE = \sqrt{\frac{1}{N} \sum_{p=1}^{N} \frac{\sum_{e=1}^{Mp} Dist(xe'^{p}_{ij}, xe^{p}_{ij})^2}{\sum_{e=1}^{Mp} (xe'^{p}_{ij})^2}} \qquad (22)$$

d) Domain distance: Hausdorff distance HD

HD measures how far two sets are from each other as in Equations 23-24 below.

$$HD(A, B) = \max\{ \sup_{a \in A}(\inf_{b \in B}(Dist(a,b))), \sup_{b \in B}(\inf_{a \in A}(Dist(a,b)) \} \qquad (23)$$

$$HD(M'^{p}_{ij}, M^{p}_{ij}) = \frac{1}{N} \sum_{p=1}^{N} \max\{ \sup_{m'^{p}_{ij} \in M'^{p}_{ij}}(\inf_{m^{p}_{ij} \in M^{p}_{ij}}(Dist(m'^{p}_{ij}, m^{p}_{ij}))), \sup_{m^{p}_{ij} \in M^{p}_{ij}}(\inf_{m'^{p}_{ij} \in M'^{p}_{ij}}(Dist(m'^{p}_{ij}, m^{p}_{ij})) \qquad (24)$$

Where
- sup() is the supremum
- inf() is the infimum
- Dist (a,b) is the distance between point a in the first set and point b in the second set.

$\inf_{b \in B}(d(a,b))$ is the infimum distance between point a and all the points in set B

$HD95$ metric replaces the supremum in the equation by the 95 percentile, which results in less sensitivity to outliers.

e) Domain distance: Center of mass COM measures the displacement between two the center of two sets A, B as shown in Equations 25 and 26

$$COM(A, B) = dist(\ Center(A), Center(B)\ ) \quad (25)$$

$$Center(A) = mean(x), \forall x \in A \quad (26)$$

6.3 Segmentation measures

a) The dice similarity coefficient (DSC) measures the overlap between two segmentation sets $L_{\emptyset j}^p, L_{ij}^p$ as in Equation 27 below.

$$DSC(L_{\emptyset j}^p, L_{ij}^p) = \frac{2\ |L_{\emptyset j}^p \cap L_{ij}^p|}{|L_{\emptyset j}^p| + |L_{ij}^p|} \quad (27)$$

DSC is equivalent to the F1 score used in classification problems, where the segmentation problem is a classification problem on the pixel level, in which a pixel/point is assigned to a segmentation label that could be true or false. F1 = 2TP/(FP+FN +2TP).

b) The Jaccard coefficient is similar to DSC with a slight modification shown in Equation 28.

$$Jaccard(L_{\emptyset j}^p, L_{ij}^p) = \frac{|L_{\emptyset j}^p \cap L_{ij}^p|}{|L_{\emptyset j}^p \cup L_{ij}^p|} \quad (28)$$

6.4 Correlation measures

a) The normalized cross-correlation nCC for two images $<X_{\emptyset i}^p, Y_{\emptyset i}^p>$ and $<X_{ji}^p, Y_{ji}^p>$ with a shared domain $X_{\emptyset i}^p = X_{ji}^p$ can be expressed in Equations 29-31 below

$$nCC(Y_{\emptyset i}^p, Y_{ji}^p) = \frac{(Y_{\emptyset i}^p - \overline{Y_{\emptyset i}^p}) \cdot (Y_{ji}^p - \overline{Y_{ji}^p})}{\sqrt{|Y_{\emptyset i}^p - \overline{Y_{\emptyset i}^p}| \times |Y_{ji}^p - \overline{Y_{ji}^p}|}} = \frac{\sum_{e=1}^{Lp}(ye_{\emptyset i}^p - \overline{Y_{\emptyset i}^p}) \times (ye_{ji}^p - \overline{Y_{ji}^p})}{\sqrt{\sum_{e=1}^{Lp}(ye_{\emptyset i}^p - \overline{Y_{\emptyset i}^p})^2 \times \sum_{e=1}^{Lp}(ye_{ji}^p - \overline{Y_{ji}^p})^2}} \text{ given that } \forall e\ xe_{\emptyset i}^p = xe_{ji}^p \quad (29)$$

$$\overline{Y} = mean(Y)\ \frac{\sum_{e=1}^{Lp} ye}{Lp} \quad (30)$$

$$|Y| = \sqrt{\sum_{e=1}^{Lp}(ye)^2} \quad (31)$$

Note: it has been reported that correlation is a better objective function than MSE, and RMSE for image registration (Zitova et al., 2003; Haskins et al., 2020).

b) The normalized local cross-correlation nLCC measures nCC among local batches (small chunks of an image), instead of full images. Let an image be divided into B local batches (subsets) such that $Y^{pb} \in Y^p$, each with a regular size of R pixels/elements.

$$nLCC(Y_{\emptyset i}^p, Y_{ji}^p) = \frac{1}{B}\sum_{b=1}^{B} nLCC(Y_{\emptyset i}^{pb}, Y_{ji}^{pb}) \quad (32)$$

6.5 The smoothness of the registration field

A non-smooth registration field can relocate a pixel far away from all its adjacent pixels after registration, however, a smooth registration field is more likely to keep nearby pixels relatively close to each other after relocation.

    a. The determinant of the Jacobian (JOCA)

$$JOCA = |\ J(\ \varphi\ )\ | \quad (33)$$

    b. The standard deviation of log Jacobian (SDlogJ)

$$SDlogJ = \sigma(\log(\ JOCA\ )\ ) \quad (34)$$

6.6 Model size

A model size can be expressed by the number of bytes that a model occupies in a storage device or the total number of its parameters.

6.7 Clinical-based evaluation

Virtual evaluation using computer-based metrics (above) may not always align perfectly with the practical evaluation by medical experts. Thus, clinical-based evaluation and involvement of domain experts from the medical field have been recommended by Chen, X., Wang et al. (2022) to characterize the reliability of MIR tools (Huang et al., 2022).

The challenges of MIR assessment included 1) the lack of ground truth labels in practical scenarios makes it difficult to evaluate an MIR outcome convincingly. 2) Medical experts' assessment could be subjective and may vary among experts. 3) Instable outcomes of some MIR algorithms, which yield different outcomes of different registration qualities for the same input image. 4) the quality of data can have a substantial impact on registration results, making it challenging to compare algorithms across datasets with varying quality (Chen, T. et al.,2022).

Deep learning in medical image registration: introduction and survey

### 7.0 Medical imaging datasets
A list of public datasets used in the literature was summarized in Table 5. The datasets were categorized based on the region of interest (ROI) such as brain, chest, …etc., and the medical imaging type.

Table 5. Medical images datasets

| ROI | Modality | Dataset | Link |
|---|---|---|---|
| Brain | MR | OASIS: Open Access Series of Imaging Studies | https://www.oasis-brains.org/ |
| | MR | LPBA40: The LONI Probabilistic Brain Atlas | https://www.loni.usc.edu/research/atlases |
| | MR | ADNI: Alzheimer's Disease Neuroimaging Initiative | https://adni.loni.usc.edu/ |
| | MR | IXI | https://brain-development.org/ixi-dataset/ |
| | MR | IBIS | |
| | MR | IBSR: The Internet Brain Segmentation Repository | https://www.nitrc.org/projects/ibsr |
| | MR | ADHD-200: Attention Deficit Hyperactivity Disorder | http://fcon_1000.projects.nitrc.org/indi/adhd200/ |
| | MR | PPMI | https://www.ppmi-info.org/access-data-specimens/download-data |
| | MR | CUMC12, MGH10 | https://www.synapse.org/#!Synapse:syn3207203 |
| | MR | ABIDE: Autism Brain Imaging Data Exchange | http://fcon_1000.projects.nitrc.org/indi/abide/ |
| | MR | BraTS2017: Brain Tumor Segmentation | https://www.med.upenn.edu/sbia/brats2017/data.html |
| | MR | Mindboggle | https://mindboggle.info/data.html |
| | MR simulated | BrainWeb | https://brainweb.bic.mni.mcgill.ca/brainweb/ |
| | MR / US | BITE: Brain Images of Tumors for Evaluation database | https://nist.mni.mcgill.ca/data/ |
| | MR / US | CuRIOUS2018 | https://curious2018.grand-challenge.org/Data/ |
| | MR / US | RESECT: a clinical database of pre-operative MRI and intra-operative ultrasound in low-grade glioma surgeries | https://archive.norstore.no/pages/public/datasetDetail.jsf?id=10.11582/2017.00004 |
| Prostate | MR | Prostate-3T | https://wiki.cancerimagingarchive.net/display/Public/Prostate-3T |
| | MR | PROMISE12: Prostate MR Image Segmentation 2012 | https://zenodo.org/record/8026660 |
| | MR | Prostate Fused-MRI-Pathology | https://wiki.cancerimagingarchive.net/pages/viewpage.action?pageId=23691514 |
| Spine | CT, MR depending on the dataset | SpineWeb library | http://spineweb.digitalimaginggroup.ca/Index.php?n=Main.Datasets |
| Knee | MR | OAI | https://nda.nih.gov/oai/ |
| Chest | CT | NLST | https://cdas.cancer.gov/datasets/nlst/ |
| | CT | SPARE | https://image-x.sydney.edu.au/spare-challenge/ |
| | XRAY | NIH ChestXray14 | https://nihcc.app.box.com/v/ChestXray-NIHCC |
| | XRAY | JSRT: Japanese Society of Radiological Technology | http://db.jsrt.or.jp/eng.php http://imgcom.jsrt.or.jp/minijsrtdb/ |
| | XRAY | Tuberculosis image datasets | https://lhncbc.nlm.nih.gov/LHC-downloads/downloads.html#tuberculosis-image-data-sets |
| Lung | CT | POPI | https://www.creatis.insa-lyon.fr/rio/popi-model_original_page |
| | CT | NLST | https://cdas.cancer.gov/datasets/nlst/ |
| | CT | SPARE | https://image-x.sydney.edu.au/spare-challenge/ |
| Heart | MR/CT | MM-WHS: Multi-Modality Whole Heart Segmentation | http://www.sdspeople.fudan.edu.cn/zhuangxiahai/0/mmwhs/ |
| | MR | SCD: The Sunnybrook Cardiac Data | https://www.cardiacatlas.org/sunnybrook-cardiac-data/ |
| Liver | CT | sliver07 | https://sliver07.grand-challenge.org/Home/ |
| Kidney | CT | KITS23 | https://kits-challenge.org/kits23/ |
| Pancreas | CT | Pancreas-CT | https://opendatalab.com/Pancreas-CT_Dataset https://wiki.cancerimagingarchive.net/display/public/pancreas-ct |
| Abdomen (kidney, liver, Spleen, Pancreas) | CT | Anatomy3 | https://visceral.eu/benchmarks/anatomy3-open/ |
| 10 ROIs | MR or CT | Medical Segmentation Decathon challenge | https://decathlon-10.grand-challenge.org/ |

### 8.0 Medical applications
Changing the frame of reference might mislead humans (like the phenomenon of not recognizing an object if it has been flipped (e.g., old/young lady face in Figure 2). Hence, it is easier for medical practitioners to evaluate a medical image in a standard reference frame (e.g., orientation, scale). Thus, registration is an essential part of medical diagnoses that depend on imaging technologies. IR was applied in retina imaging (Ho et al., 2021), breast imaging (Ringel et al., 2022; Ying et al., 2022), HIFU treatment of heart arrythmias (Dahman et al., 2022), and cross-staining alignment (Wang et al., 2022). Selected applications of MIR are discussed below.

8.1 Image-guided surgery

Image-guided surgery (IGS) incorporates imaging modalities such as CT, and US to assist surgeons during surgical procedures. For example, surgeons can visualize internal anatomy, pinpoint the location of tumors or lesions, and determine optimal incision points. Image-guided surgery enables surgeons to precisely target specific areas, and avoid critical structures during a procedure.

Before an IGS, a patient's preoperative images are loaded into a software or surgical navigation system (Wang, D. et al., 2022). The collected images are then aligned with images taken during the surgery (inter-operative) using registration algorithms. Having images with key points/landmarks improves the registration process in terms of speed and precision. The landmarks can be selected manually by medical experts on computer software (Schmidt et al., 2022; Wang, Y. et al., 2022), or they could be fiducial markers, which are small devices placed in a patient's body such as the injection of gold seeds to mark a tumor before radiation therapy. The number of landmarks needed for a precise registration can be reduced by the integration of semantic segmentation in addition to the use of a standard template (atlas) instead of preoperative images as shown by (Su et al., 2021). An alignment with no landmarks was tested by (Robertson et al., 2022) for catheter placement in non-immobilized patients.

To mention some examples of the use of MIR for IGS (Vijayan et al., 2021; Upendra et al., 2021, February), 2D inter-operative and 3D preoperative images were aligned in real-time surgical navigation systems (Ashfaq et al., 2022). A similar alignment of 2D-3D was needed for the deep brain stimulation procedure which involves the placement of neuro-electrodes into the brain to treat movement disorders such as Parkinson, and Dystonia (Uneri et al., 2021). A real-time biopsy navigation system was developed by (Dupuy et al., 2021) to align 2D US inter-operative images with 3D TRUS preoperative images and to estimate in real-time the biopsy target of a prostate based on its previous trajectory.

8.2 Tumor diagnosis and therapy

A tumor is an abnormal mass or growth of cells in the body. Tumors can develop in various tissues or organs and can be either benign or malignant. Benign tumors are non-cancerous and typically do not invade nearby tissues or spread to other parts of the body. Benign tumors are generally not life-threatening, but medical attention and/or treatment are still required. Malignant tumors, on the other side, are cancerous. They have the potential to invade surrounding tissues and can spread to other parts of the body through the bloodstream or lymphatic system. Malignant tumors grow rapidly and can be life-threatening. Medical experts often diagnose a tumor and plan therapy depending on the tumor's growth over time as recorded in aligned medical images. Accordingly, MIR has been used for radiotherapy (Fu et al., 2022; Vargas-Bedoya et al., 2022) and proton therapy (Hirotaki et al., 2022).

8.3 Motion processing

The human body experience normal deformation over time, some deformations occur at a slower pace such as the growth of bones over a lifetime (e.g., a human height grows from afew feet in newborns to several feet in adults) while some deformations occur at a faster pace such as heartbeats. The heart experiences alternating contractions and relaxations while pumping blood at a frequency of 1-3 beats per second. MIR helps to analyze such temporospatial deformations and resulting movements.

Cardiac motion was tracked by (Ye et al., 2021) using tagging magnetic resonance imaging (t-MRI), where an unsupervised bidirectional MIR model estimated the motion field between consecutive frames. (Upendra et al., 2021, November) focused on motion extraction from 4D cardiac CMRI (Cine Magnetic Resonance Imaging), mainly the development of patient-specific right ventricle (RV) models based on kinematic analysis. A DL deformable MIR was used to estimate the motion of the RV and generate isosurface meshes of cardiac geometry.

Respiratory movement can affect the quality of medical imaging by causing motion blur. To overcome this (Hou et al., 2022) proposed an unsupervised MIR framework for respiratory motion correction in PET (Positron Emission Tomography) images. (Chaudhary et al., 2022) focused on lung tissue expansion which is typically estimated by registering multiple scans. To reduce the number of needed scans, Chaudhary et al., (2022) proposed the use of generative adversarial learning to estimate local tissue expansion of lungs from a single CT scan.

2D-3D motion registration of bones was addressed in (Djurabekova et al., 2022) by manipulating segmented bones from static scans and matching digitally reconstructed radiographs to X-ray projections. The bones were, particularly foot and ankle structure.

## 9.0 Other research directions

### 9.1 Transformers

Transformers are a DL architecture that uses the attention mechanism solely dispensing with conventional and recurrent units (Vaswani et al., 2017). Transformers have contributed to noticeable improvements in computer vision, audio processing, and language processing tasks (Lin et al, 2022). The improvement can be seen in products like GPT-2, and ChatGPT which are examples of Generative Pre-trained Transformers (GPT). Transformers can be decomposed into basic/abstract mathematical components that distinguished them from recurrent and convolutional networks: 1) the position encoding, which explicitly feeds the position of a token as an input, 2) the product operation between features which is manifested explicitly in the product between the key and the query of the attention mechanism, and implicitly within the exponential function of the SoftMax ($e^{a+b} = e^a \times e^b$). 3) the exponential function which represents a transformation into another space.

In MIR, (Mok et al., 2022) proposed the use of the attention mechanism for affine MIR such that multi-head attention was used in the encoder, and convolutional units in the decoder. Transformers were embedded partially for deformable MIR in Transmorph (Chen, J. et al., 2022). Transmorph is a coarse-fine IR such that affine alignment is conducted in the first stage followed by deformable alignment in the second stage. The latter stage is a Voxelmorph-like registration with U-Net architecture except that the encoder part consists of transformers instead of ConvNets. Transmorph introduced transformers (self-attention blocks) as a part of the encoder only but not the decoder. Ma et al. (2022) attributed the difficulty of developing transformers for MIR to the large number of trainable parameters of a transformer unit compared to convolutional units. To reduce the number of parameters, the authors proposed the use of both convolution units and transformer units in an MIR model - SymTrans (Ma et al., 2022). SymTrans embedded transformers in both the encoder and the decoder (2 blocks in the encoder and 2 in the decoder).

The utilization of transformers in MIR was not as fast and revolutionary as it was in other domains. That could be attributed to the relatively small number of images in MIR datasets compared to other tasks. For example, millions of images were used for the ViLT model (Kim et al., 2021), and up to 0.8 billion images for the GiT model (Wang, J. et al., 2022).

### 9.2 No Registration

Another potential research direction is the elimination of the image registration step from the medical image analysis pipeline. In theory, an end-to-end deep learning model learns an automatic medical image analysis task (e.g., disease detection) without an explicit registration step. In (Chen, X., Zhang, et al, 2022), the authors proposed the elimination of the registration step entirely by the development of a breast cancer prediction model using vision transformers and multi-view images.

9.3 Other research directions explored before include Fourier transform-based IR (Zitova et al., 2003), Reinforcement learning based IR (Chen, X. et al., 2021; George et al., 2021; Sutton et al., 1994), and GANs-based MIR (Xiao et al., 2021; Chaudhary et al., 2022; Dey et al., 2021; Goodfellow et al., 2020). There could be further research interest in the mentioned MIR research directions in the future.

**Acknowledgment**

This work was funded partially by Service Public de Wallonie Recherche under grant n°2010235 ARIAC by digitalwallonia4.ai

## References

Abbasi, S., Tavakoli, M., Boveiri, H. R., Shirazi, M. A. M., Khayami, R., Khorasani, H., ... & Mehdizadeh, A. (2022). Medical image registration using unsupervised deep neural network: A scoping literature review. Biomedical Signal Processing and Control, 73, 103444.

Adelson, E. H., Anderson, C. H., Bergen, J. R., Burt, P. J., & Ogden, J. M. (1984). Pyramid methods in image processing. RCA engineer, 29(6), 33-41.

Alcaín, E., Fernández, P. R., Nieto, R., Montemayor, A. S., Vilas, J., Galiana-Bordera, A., ... & Torrado-Carvajal, A. (2021). Hardware architectures for real-time medical imaging. Electronics, 10(24), 3118.

Andreadis, G., Bosman, P. A., & Alderliesten, T. (2022, April). Multi-objective dual simplex-mesh based deformable image registration for 3D medical images-proof of concept. In Medical Imaging 2022: Image Processing (Vol. 12032, pp. 744-750). SPIE.

# Deep learning in medical image registration: introduction and survey

Arun, K. S., Huang, T. S., & Blostein, S. D. (1987). Least-squares fitting of two 3-D point sets. IEEE Transactions on pattern analysis and machine intelligence, (5), 698-700.

Ashfaq, M., Minallah, N., Frnda, J., & Behan, L. (2022). Multi-Modal Rigid Image Registration and Segmentation Using Multi-Stage Forward Path Regenerative Genetic Algorithm. Symmetry, 14(8), 1506.

Avants, B. B., Epstein, C. L., Grossman, M., & Gee, J. C. (2008). Symmetric diffeomorphic image registration with cross-correlation: evaluating automated labeling of elderly and neurodegenerative brain. Medical image analysis, 12(1), 26-41.

Avants, B. B., Tustison, N. J., Stauffer, M., Song, G., Wu, B., & Gee, J. C. (2014). The Insight ToolKit image registration framework. Frontiers in neuroinformatics, 8, 44.

Balakrishnan, G., Zhao, A., Sabuncu, M. R., Guttag, J., & Dalca, A. V. (2019). VoxelMorph: a learning framework for deformable medical image registration. IEEE transactions on medical imaging, 38(8), 1788-1800.

Ban, Y., Wang, Y., Liu, S., Yang, B., Liu, M., Yin, L., & Zheng, W. (2022). 2D/3D multimode medical image alignment based on spatial histograms. Applied Sciences, 12(16), 8261.

Bashkanov, O., Meyer, A., Schindele, D., Schostak, M., Tönnies, K. D., Hansen, C., & Rak, M. (2021, April). Learning Multi-Modal Volumetric Prostate Registration With Weak Inter-Subject Spatial Correspondence. In 2021 IEEE 18th International Symposium on Biomedical Imaging (ISBI) (pp. 1817-1821). IEEE.

Begum, N., Badshah, N., Rada, L., Ademaj, A., Ashfaq, M., & Atta, H. (2022). An improved multi-modal joint segmentation and registration model based on Bhattacharyya distance measure. Alexandria Engineering Journal, 61(12), 12353-12365.

Bengio, Y., Louradour, J., Collobert, R., & Weston, J. (2009, June). Curriculum learning. In Proceedings of the 26th annual international conference on machine learning (pp. 41-48).

Bouaziz, S., Tagliasacchi, A., & Pauly, M. (2013, August). Sparse iterative closest point. In Computer graphics forum (Vol. 32, No. 5, pp. 113-123). Oxford, UK: Blackwell Publishing Ltd.

Bronstein, M. M., Bruna, J., Cohen, T., & Veličković, P. (2021). Geometric deep learning: Grids, groups, graphs, geodesics, and gauges. arXiv preprint arXiv:2104.13478.

Brown, L. G. (1992). A survey of image registration techniques. ACM computing surveys (CSUR), 24(4), 325-376.

Burduja, M., & Ionescu, R. T. (2021, September). Unsupervised medical image alignment with curriculum learning. In 2021 IEEE International Conference on Image Processing (ICIP) (pp. 3787-3791). IEEE.

Chaudhary, M. F., Gerard, S. E., Wang, D., Christensen, G. E., Cooper, C. B., Schroeder, J. D., ... & Reinhardt, J. M. (2022, March). Single volume lung biomechanics from chest computed tomography using a mode preserving generative adversarial network. In 2022 IEEE 19th International Symposium on Biomedical Imaging (ISBI) (pp. 1-5). IEEE.

Chen, J., Frey, E. C., He, Y., Segars, W. P., Li, Y., & Du, Y. (2022). Transmorph: Transformer for unsupervised medical image registration. Medical image analysis, 82, 102615.

Chen, T., Yuan, M., Tang, J., & Lu, L. (2022). Digital Analysis of Smart Registration Methods for Magnetic Resonance Images in Public Healthcare. Frontiers in Public Health, 10, 896967.

Chen, X., Diaz-Pinto, A., Ravikumar, N., & Frangi, A. F. (2021). Deep learning in medical image registration. Progress in Biomedical Engineering, 3(1), 012003.

Chen, X., Wang, X., Zhang, K., Fung, K. M., Thai, T. C., Moore, K., ... & Qiu, Y. (2022). Recent advances and clinical applications of deep learning in medical image analysis. Medical Image Analysis, 79, 102444.

Chen, X., Zhang, K., Abdoli, N., Gilley, P. W., Wang, X., Liu, H., ... & Qiu, Y. (2022). Transformers improve breast cancer diagnosis from unregistered multi-view mammograms. Diagnostics, 12(7), 1549.

Cooper, L. A., "Mental rotation of random two-dimensional shapes," Cognitive psychology, 7(1), 20-43(1975).

Dahman, B., Bessier, F., & Dillenseger, J. L. (2022, April). Ultrasound to CT rigid image registration using CNN for the HIFU treatment of heart arrhythmias. In Medical Imaging 2022: Image-Guided Procedures, Robotic Interventions, and Modeling (Vol. 12034, pp. 246-252). SPIE.

Decuyper, M., Maebe, J., Van Holen, R., & Vandenberghe, S. (2021). Artificial intelligence with deep learning in nuclear medicine and radiology. EJNMMI physics, 8(1), 81.

Denton, E. L., Chintala, S., & Fergus, R. (2015). Deep generative image models using a laplacian pyramid of adversarial networks. Advances in neural information processing systems, 28.

Dey, N., Ren, M., Dalca, A. V., & Gerig, G. (2021). Generative adversarial registration for improved conditional deformable templates. In Proceedings of the IEEE/CVF international conference on computer vision (pp. 3929-3941).

Dida, H., Charif, F., & Benchabane, A. (2022). Image registration of computed tomography of lung infected with COVID-19 using an improved sine cosine algorithm. Medical & Biological Engineering & Computing, 60(9), 2521-2535.

# Deep learning in medical image registration: introduction and survey

Ding, W., Li, L., Zhuang, X., & Huang, L. (2022). Cross-Modality Multi-Atlas Segmentation via Deep Registration and Label Fusion. IEEE Journal of Biomedical and Health Informatics, 26(7), 3104-3115.

Ding, Z., & Niethammer, M. (2022). Aladdin: Joint atlas building and diffeomorphic registration learning with pairwise alignment. In Proceedings of the IEEE/CVF conference on computer vision and pattern recognition (pp. 20784-20793).

Djurabekova, N., Goldberg, A., Hawkes, D., Long, G., Lucka, F., & Betcke, M. M. (2022, October). 2D-3D motion registration of rigid objects within a soft tissue structure. In 7th International Conference on Image Formation in X-Ray Computed Tomography (Vol. 12304, pp. 518-526). SPIE.

Dossun, C., Niederst, C., Noel, G., & Meyer, P. (2022). Evaluation of DIR algorithm performance in real patients for radiotherapy treatments: A systematic review of operator-dependent strategies. Physica Medica, 101, 137-157.

Dupuy, T., Beitone, C., Troccaz, J., & Voros, S. (2021, February). 2D/3D deep registration for real-time prostate biopsy navigation. In Medical Imaging 2021: Image-Guided Procedures, Robotic Interventions, and Modeling (Vol. 11598, pp. 463-471). SPIE.

Estépar, R. S. J., Brun, A., & Westin, C. F. (2004). Robust generalized total least squares iterative closest point registration. In Medical Image Computing and Computer-Assisted Intervention–MICCAI 2004: 7th International Conference, Saint-Malo, France, September 26-29, 2004. Proceedings, Part I 7 (pp. 234-241). Springer Berlin Heidelberg.

Fitzpatrick, J. M., Hill, D. L., & Maurer, C. R. (2000). Image registration. Handbook of medical imaging, 2, 447-513.

Fu, H. J., Chen, P. Y., Yang, H. Y., Tsang, Y. W., & Lee, C. Y. (2022). Liver-directed stereotactic body radiotherapy can be reliably delivered to selected patients without internal fiducial markers—A case series. Journal of the Chinese Medical Association, 85(10), 1028-1032.

Ganeshaaraj, G., Kaushalya, S., Kondarage, A. I., Karunaratne, A., Jones, J. R., & Nanayakkara, N. D. (2022, July). Semantic Segmentation of Micro-CT Images to Analyze Bone Ingrowth into Biodegradable Scaffolds. In 2022 44th Annual International Conference of the IEEE Engineering in Medicine & Biology Society (EMBC) (pp. 3830-3833). IEEE.

Gao, X., & Zheng, G. (2022, August). ACSGRegNet: A Deep Learning-based Framework for Unsupervised Joint Affine and Diffeomorphic Registration of Lumbar Spine CT via Cross-and Self-Attention Fusion. In Proceedings of the 2022 International Conference on Intelligent Medicine and Health (pp. 57-63).

George, Y., Sedai, S., Antony, B. J., Ishikawa, H., Wollstein, G., Schuman, J. S., & Garnavi, R. (2021, April). Dueling deep Q-network for unsupervised inter-frame eye movement correction in optical coherence tomography volumes. In 2021 IEEE 18th International Symposium on Biomedical Imaging (ISBI) (pp. 1595-1599). IEEE.

Goodfellow, I., Pouget-Abadie, J., Mirza, M., Xu, B., Warde-Farley, D., Ozair, S., ... & Bengio, Y. (2020). Generative adversarial networks. Communications of the ACM, 63(11), 139-144.

Goyal, A., & Bengio, Y. (2022). Inductive biases for deep learning of higher-level cognition. Proceedings of the Royal Society A, 478(2266), 20210068.

Haskins, G., Kruger, U., & Yan, P. (2020). Deep learning in medical image registration: a survey. Machine Vision and Applications, 31, 1-18.

Himthani, N., Brunn, M., Kim, J. Y., Schulte, M., Mang, A., & Biros, G. (2022). CLAIRE—Parallelized Diffeomorphic Image Registration for Large-Scale Biomedical Imaging Applications. Journal of Imaging, 8(9), 251.

Hirotaki, K., Moriya, S., Akita, T., Yokoyama, K., & Sakae, T. (2022). Image preprocessing to improve the accuracy and robustness of mutual-information-based automatic image registration in proton therapy. Physica Medica, 101, 95-103.

Ho, C. J., Wang, Y., Zhang, J., Nguyen, T., & An, C. (2021, October). A Convolutional Neural Network Pipeline For Multi-Temporal Retinal Image Registration. In 2021 18th International SoC Design Conference (ISOCC) (pp. 27-28). IEEE.

Hoffmann, M., Billot, B., Greve, D. N., Iglesias, J. E., Fischl, B., & Dalca, A. V. (2021). SynthMorph: learning contrast-invariant registration without acquired images. IEEE transactions on medical imaging, 41(3), 543-558.

Hou, Y., He, J., & She, B. (2022). Respiratory Motion Correction on PET Images Based on 3D Convolutional Neural Network. KSII Transactions on Internet and Information Systems (TIIS), 16(7), 2191-2208.

Huang, J., Shlobin, N. A., Lam, S. K., & DeCuypere, M. (2022). Artificial intelligence applications in pediatric brain tumor imaging: A systematic review. World neurosurgery, 157, 99-105.

Kasban, H., El-Bendary, M. A. M., & Salama, D. H. (2015). A comparative study of medical imaging techniques. International Journal of Information Science and Intelligent System, 4(2), 37-58.

Kim, W., Son, B., & Kim, I. (2021, July). Vilt: Vision-and-language transformer without convolution or region supervision. In International Conference on Machine Learning (pp. 5583-5594). PMLR.

Kujur, S. S., & Sahana, S. K. (2022). Medical image registration utilizing tissue P systems. Frontiers in Pharmacology, 13, 949872.

Lai, W. S., Huang, J. B., Ahuja, N., & Yang, M. H. (2017). Deep laplacian pyramid networks for fast and accurate super-resolution. In Proceedings of the IEEE conference on computer vision and pattern recognition (pp. 624-632).

# Deep learning in medical image registration: introduction and survey

Lee, H. H., Tang, Y., Bao, S., Xu, Y., Yang, Q., Yu, X., ... & Landman, B. A. (2022, April). Supervised deep generation of high-resolution arterial phase computed tomography kidney substructure atlas. In Medical Imaging 2022: Image Processing (Vol. 12032, pp. 736-743). SPIE.

Li, Y. X., Tang, H., Wang, W., Zhang, X. F., & Qu, H. (2022). Dual attention network for unsupervised medical image registration based on VoxelMorph. Scientific Reports, 12(1), 16250.

Lin, T., Wang, Y., Liu, X., & Qiu, X. (2022). A survey of transformers. AI Open.

Liu, P., Wang, F., Teodoro, G., & Kong, J. (2021, April). Histopathology image registration by integrated texture and spatial proximity based landmark selection and modification. In 2021 IEEE 18th International Symposium on Biomedical Imaging (ISBI) (pp. 1827-1830). IEEE.

Ma, M., Xu, Y., Song, L., & Liu, G. (2022). Symmetric transformer-based network for unsupervised image registration. Knowledge-Based Systems, 257, 109959.

Maillard, M., François, A., Glaunès, J., Bloch, I., & Gori, P. (2022, March). A deep residual learning implementation of metamorphosis. In 2022 IEEE 19th International Symposium on Biomedical Imaging (ISBI) (pp. 1-4). IEEE.

Meng, M., Bi, L., Fulham, M., Feng, D. D., & Kim, J. (2022). Enhancing medical image registration via appearance adjustment networks. NeuroImage, 259, 119444.

Mitchell, T. M. (1980). The need for biases in learning generalizations (pp. 184-191). New Jersey: Department of Computer Science, Laboratory for Computer Science Research, Rutgers Univ.

Mitchell, W. J. (1984). What is an Image? New literary history, 15(3), 503-537.

Modat, M., Ridgway, G. R., Taylor, Z. A., Lehmann, M., Barnes, J., Hawkes, D. J., ... & Ourselin, S. (2010). Fast free-form deformation using graphics processing units. Computer methods and programs in biomedicine, 98(3), 278-284.

Modersitzki, J. (2003). Numerical methods for image registration. OUP Oxford.

Mok, T. C., & Chung, A. (2022). Affine medical image registration with coarse-to-fine vision transformer. In Proceedings of the IEEE/CVF Conference on Computer Vision and Pattern Recognition (pp. 20835-20844).

Naik, R. R., Anitha, H., Bhat, S. N., Ampar, N., & Kundangar, R. (2022). Realistic C-arm to pCT registration for vertebral localization in spine surgery: A hybrid 3D-2D registration framework for intraoperative vertebral pose estimation. Medical & Biological Engineering & Computing, 60(8), 2271-2289.

Nazib, A., Fookes, C., Salvado, O., & Perrin, D. (2021, April). A multiple decoder CNN for inverse consistent 3D image registration. In 2021 IEEE 18th International Symposium on Biomedical Imaging (ISBI) (pp. 904-907). IEEE.

Ourselin, S., Roche, A., Subsol, G., Pennec, X., & Ayache, N. (2001). Reconstructing a 3D structure from serial histological sections. Image and vision computing, 19(1-2), 25-31.

Oyallon, E., & Mallat, S. (2015). Deep roto-translation scattering for object classification. In Proceedings of the IEEE Conference on Computer Vision and Pattern Recognition (pp. 2865-2873).

Park, H., Lee, G. M., Kim, S., Ryu, G. H., Jeong, A., Sagong, M., & Park, S. H. (2022, March). A meta-learning approach for medical image registration. In 2022 IEEE 19th International Symposium on Biomedical Imaging (ISBI) (pp. 1-5). IEEE.

Ringel, M. J., Richey, W. L., Heiselman, J., Luo, M., Meszoely, I. M., & Miga, M. I. (2022, April). Breast image registration for surgery: insights on material mechanics modeling. In Medical Imaging 2022: Image-Guided Procedures, Robotic Interventions, and Modeling (Vol. 12034, pp. 223-230). SPIE.

Robertson, F. C., Raahil, M. S., Amich, J. M., Essayed, W. I., Lal, A., Lee, B. H., ... & Gormley, W. B. (2021). Frameless neuronavigation with computer vision and real-time tracking for bedside external ventricular drain placement: a cadaveric study. Journal of Neurosurgery, 136(5), 1475-1484.

Saadat, S., Perriman, D., Scarvell, J. M., Smith, P. N., Galvin, C. R., Lynch, J., & Pickering, M. R. (2022). An efficient hybrid method for 3D to 2D medical image registration. International Journal of Computer Assisted Radiology and Surgery, 17(7), 1313-1320.

Saiti, E., & Theoharis, T. (2022). Multimodal registration across 3D point clouds and CT-volumes. Computers & Graphics, 106, 259-266.

Santarossa, M., Tatli, A., von der Burchard, C., Andresen, J., Roider, J., Handels, H., & Koch, R. (2022). Chronological Registration of OCT and Autofluorescence Findings in CSCR: Two Distinct Patterns in Disease Course. Diagnostics, 12(8), 1780.

Schmidt, C., & Overhoff, H. M. (2023). Impact of PCA-based preprocessing and different CNN structures on deformable registration of sonograms. arXiv preprint arXiv:2301.08802.

Stewart, C., Ibanez, L. (2004). Image Registration Course. Rensselaer Polytechnic Institute

Stallings, R. (2010, April 22). Printing Lingo: What does Registration mean? Formax printing. https://www.formaxprinting.com/blog/2010/04/printing-lingo-what-does-registration-mean

# Deep learning in medical image registration: introduction and survey

Su, S., Song, G., & Zhao, Y. (2021, November). 3D Registration of the Point Cloud Data Using Parameter Adaptive Super4PCS Algorithm in Medical Image Analysis. In 2021 4th International Conference on Digital Medicine and Image Processing (pp. 1-6).

Sutton, R. S., & Barto, A. G. (1998). Introduction to reinforcement learning (Vol. 135, pp. 223-260). Cambridge: MIT Press.

Talwar, V., Chufal, K. S., & Joga, S. (2021). Artificial Intelligence: A New Tool in Oncologist's Armamentarium. Indian Journal of Medical and Paediatric Oncology, 42(06), 511-517.

Terpstra, M. L., Maspero, M., Sbrizzi, A., & van den Berg, C. A. (2022). ⊥-loss: A symmetric loss function for magnetic resonance imaging reconstruction and image registration with deep learning. Medical Image Analysis, 80, 102509.

Thirion, J. P. (1996, June). Non-rigid matching using demons. In Proceedings CVPR IEEE Computer Society Conference on Computer Vision and Pattern Recognition (pp. 245-251). IEEE.

Uneri, A., Wu, P., Jones, C. K., Ketcha, M. D., Vagdargi, P., Han, R., ... & Siewerdsen, J. H. (2021, February). Data-driven deformable 3D-2D registration for guiding neuroelectrode placement in deep brain stimulation. In Medical Imaging 2021: Image-Guided Procedures, Robotic Interventions, and Modeling (Vol. 11598, pp. 349-353). SPIE.

Upendra, R. R., Hasan, S. K., Simon, R., Wentz, B. J., Shontz, S. M., Sacks, M. S., & Linte, C. A. (2021, November). Motion extraction of the right ventricle from 4D cardiac cine MRI using a deep learning-based deformable registration framework. In 2021 43rd Annual International Conference of the IEEE Engineering in Medicine & Biology Society (EMBC) (pp. 3795-3799). IEEE.

Upendra, R. R., Simon, R., & Linte, C. A. (2021, February). Joint deep learning framework for image registration and segmentation of late gadolinium enhanced MRI and cine cardiac MRI. In Medical Imaging 2021: Image-Guided Procedures, Robotic Interventions, and Modeling (Vol. 11598, pp. 96-103). SPIE.

Van Houtte, J., Audenaert, E., Zheng, G., & Sijbers, J. (2022). Deep learning-based 2D/3D registration of an atlas to biplanar X-ray images. International Journal of Computer Assisted Radiology and Surgery, 17(7), 1333-1342.

Vargas-Bedoya, E., Rivera, J. C., Puerta, M. E., Angulo, A., Wahl, N., & Cabal, G. (2022). Contour Propagation for Radiotherapy Treatment Planning Using Nonrigid Registration and Parameter Optimization: Case Studies in Liver and Breast Cancer. Applied Sciences, 12(17), 8523.

Vaswani, A., Shazeer, N., Parmar, N., Uszkoreit, J., Jones, L., Gomez, A. N., ... & Polosukhin, I. (2017). Attention is all you need. Advances in neural information processing systems, 30.

Vijayan, R. C., Han, R., Wu, P., Sheth, N. M., Vagdargi, P., Vogt, S., ... & Uneri, A. (2021, February). Fluoroscopic guidance of a surgical robot: pre-clinical evaluation in pelvic guidewire placement. In Medical Imaging 2021: Image-Guided Procedures, Robotic Interventions, and Modeling (Vol. 11598, pp. 393-399). SPIE.

Wang, C. W., Lee, Y. C., Khalil, M. A., Lin, K. Y., Yu, C. P., & Lien, H. C. (2022). Fast cross-staining alignment of gigapixel whole slide images with application to prostate cancer and breast cancer analysis. Scientific Reports, 12(1), 11623.

Wang, J., Yang, Z., Hu, X., Li, L., Lin, K., Gan, Z., ... & Wang, L. (2022). Git: A generative image-to-text transformer for vision and language. arXiv preprint arXiv:2205.14100.

Wang, H. J., Lee, C. Y., Lai, J. H., Chang, Y. C., & Chen, C. M. (2022). Image registration method using representative feature detection and iterative coherent spatial mapping for infrared medical images with flat regions. Scientific Reports, 12(1), 7932.

Wang, Y.; Liu, Z.; Cui, L. (2022). Research on Low Overlap Point Cloud Registration Algorithms in Medical Surgery. In Advances in Transdisciplinary Engineering. Volume 24: Advances in Machinery, Materials Science and Engineering Application. (pp. 694 - 701).

Wang, Z., Xin, J., Shen, H., Chen, Q., Wang, Z., & Wang, X. (2022). Custom 3D fMRI Registration Template Construction Method Based on Time-Series Fusion. Diagnostics, 12(8), 2013.

Wikipedia. 2023. "Printing registration." Wikimedia Foundation. Last modified Jan 25, 2023. https://en.wikipedia.org/wiki/Printing_registration.

Wu, Y., Jiahao, T. Z., Wang, J., Yushkevich, P. A., Hsieh, M. A., & Gee, J. C. (2022). Nodeo: A neural ordinary differential equation based optimization framework for deformable image registration. In Proceedings of the IEEE/CVF Conference on Computer Vision and Pattern Recognition (pp. 20804-20813).

Xiao, H., Teng, X., Liu, C., Li, T., Ren, G., Yang, R., ... & Cai, J. (2021). A review of deep learning-based three-dimensional medical image registration methods. Quantitative Imaging in Medicine and Surgery, 11(12), 4895.

Xu, Z., Yan, J., Luo, J., Wells, W., Li, X., & Jagadeesan, J. (2021, April). Unimodal cyclic regularization for training multimodal image registration networks. In 2021 IEEE 18th International Symposium on Biomedical Imaging (ISBI) (pp. 1660-1664). IEEE.

Yan, K., Cai, J., Jin, D., Miao, S., Guo, D., Harrison, A. P., ... & Lu, L. (2022). SAM: Self-supervised learning of pixel-wise anatomical embeddings in radiological images. IEEE Transactions on Medical Imaging, 41(10), 2658-2669.

Yang, Q., Atkinson, D., Fu, Y., Syer, T., Yan, W., Punwani, S., ... & Hu, Y. (2022). Cross-Modality Image Registration Using a Training-Time Privileged Third Modality. IEEE Transactions on Medical Imaging, 41(11), 3421-3431.

Deep learning in medical image registration: introduction and survey

Yang, Q., Vercauteren, T., Fu, Y., Giganti, F., Ghavami, N., Stavrinides, V., ... & Hu, Y. (2021, April). Morphological change forecasting for prostate glands using feature-based registration and Kernel density extrapolation. In 2021 IEEE 18th International Symposium on Biomedical Imaging (ISBI) (pp. 1072-1076). IEEE.

Yang, Y., Wang, F., & Guan, C. (2021). Application of Algorithm with Novel Successive Approximate Registration Method of Medical Image. In Journal of Physics: Conference Series (Vol. 1746, No. 1, p. 012043). IOP Publishing.

Ye, M., Kanski, M., Yang, D., Chang, Q., Yan, Z., Huang, Q., ... & Metaxas, D. (2021). Deeptag: An unsupervised deep learning method for motion tracking on cardiac tagging magnetic resonance images. In Proceedings of the IEEE/CVF conference on computer vision and pattern recogni

Ying, J., Cattell, R., Zhao, T., Lei, L., Jiang, Z., Hussain, S. M., ... & Huang, C. (2022). Two fully automated data-driven 3D whole-breast segmentation strategies in MRI for MR-based breast density using image registration and U-Net with a focus on reproducibility. Visual Computing for Industry, Biomedicine, and Art, 5(1), 25.

Zhang, J., Liu, F., Yu, X., Ma, Y., & Zhao, X. (2021, June). A 3D medical image registration method based on multi-scale feature fusion. In Journal of Physics: Conference Series (Vol. 1948, No. 1, p. 012057). IOP Publishing.

Zhang, L., Zhou, L., Li, R., Wang, X., Han, B., & Liao, H. (2021, April). Cascaded feature warping network for unsupervised medical image registration. In 2021 IEEE 18th International Symposium on Biomedical Imaging (ISBI) (pp. 913-916). IEEE.

Zhang, Y. N., Xia, K. R., Li, C. Y., Wei, B. L., & Zhang, B. (2021). Review of breast cancer pathologigcal image processing. BioMed research international, 2021, 1-7.tion (pp. 7261-7271).

Zhu, W., Huang, Y., Xu, D., Qian, Z., Fan, W., & Xie, X. (2021, May). Test-time training for deformable multi-scale image registration. In 2021 IEEE International Conference on Robotics and Automation (ICRA) (pp. 13618-13625). IEEE.

Zitova, B., & Flusser, J. (2003). Image registration methods: a survey. Image and vision computing, 21(11), 977-1000.

# Appendix A: Definitions

| | |
|---|---|
| **Affine transformation** | a geometric transformation that preserves parallelism and lines, but not necessarily angles and Euclidean distances. |
| **Automorphism** | An Isomorphism from a structure to itself. |
| **Bidirectional transformation** | $T_{i\leftrightarrow k}$ maps spaces in both directions from i to k and vice versa. |
| **Diffeomorphism** | An isomorphism of manifolds that is invertible and differentiable. |
| **Geometric deep learning** | unified geometric principles that provide a framework to study neural network architectures and that also incorporate prior physical knowledge into neural networks. |
| **Invariance** | A property that an outcome remains unchained after transformations. For example, the area of an object is invariant to rigid transformations (rotation, translation). |
| **Invertibility** | $T_{ij}$ is an invertible transformation if $\exists\ T_{ij}^{-1}$ |
| **Isomorphism** | is a mapping that preserves the structure and can be inverted such as a 1:1 correspondence between two sets. |
| **Isosurface** | a 3D surface representation of points with equal values in a 3D data distribution. |
| **Metamorphism** | the correspondence between a fixed image and a moving image is not 1:1. |
| **Multimodal registration** | input images are of different modality kinds. |
| **Orthogonal transformation** | A linear transformation preserves the inner product such that for a transformation T applied to vectors a and b, the inner product of the newly transformed vector <a',b'> = <a,b>. |
| **Rigid transformation** | is an affine transformation that preserves distances. |
| **Structure** | a set with features (e.g., operations). |
| **Symmetric Invertibility** | $\exists k \in space: T_{ij} = Tkj \circ Tik, where\ Tik = T_{jk}^{-1}$ |
| **Unimodal registration** | input images are of the same modality |

Deep learning in medical image registration: introduction and survey

## Appendix B: Statistics in figures

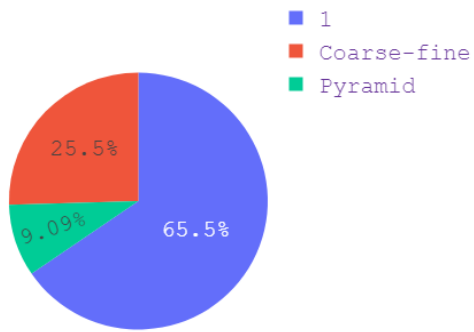

Figure 29. Multistage registration statistics

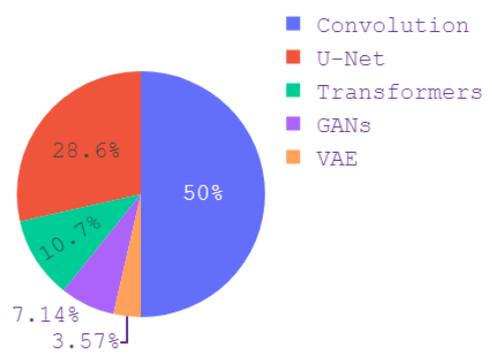

Figure 30. DL architecture statistics

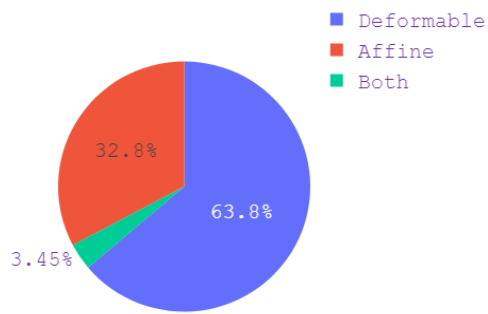

Figure 31. Deformability statistics

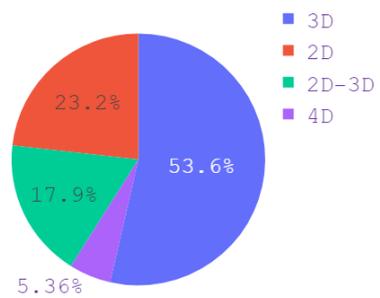

Figure 32. Dimensionality statistics

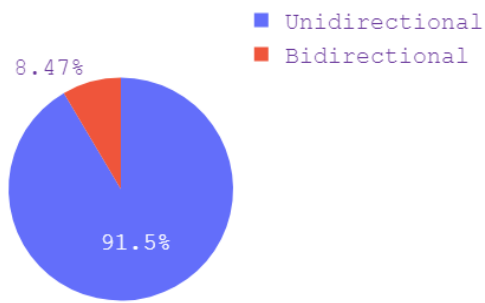
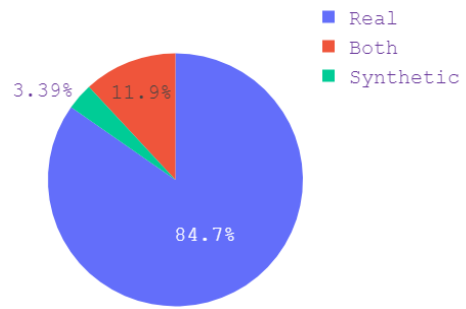

Figure 33. Invertibility statistics　　　　　　　　Figure 34. Data type statistics

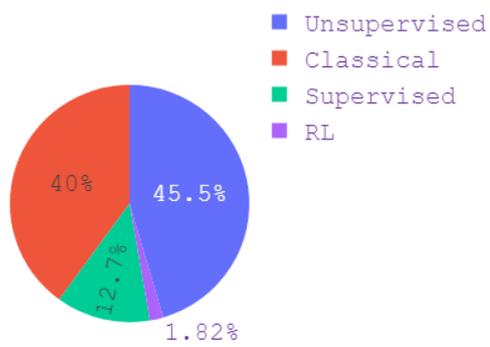
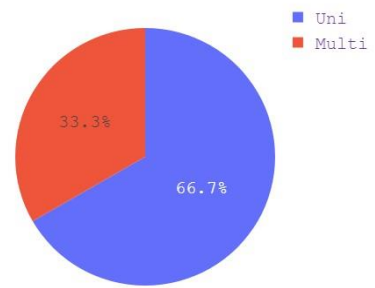

Figure 35. IR approach statistics　　　　　　　　Figure 36. Modality statistics

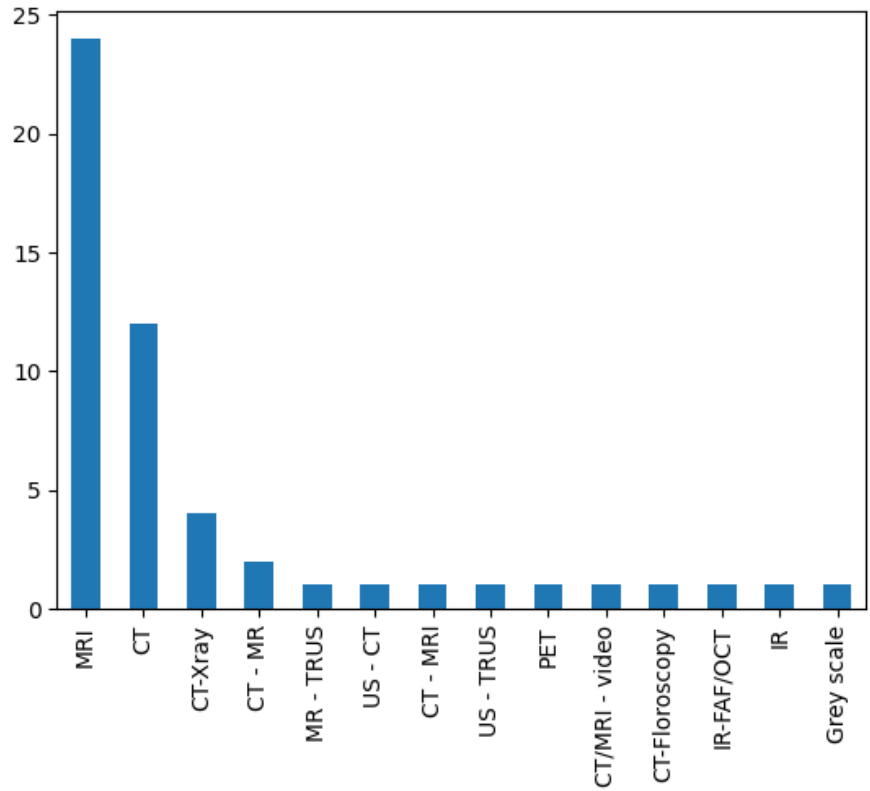

Figure 37. Modals statistics

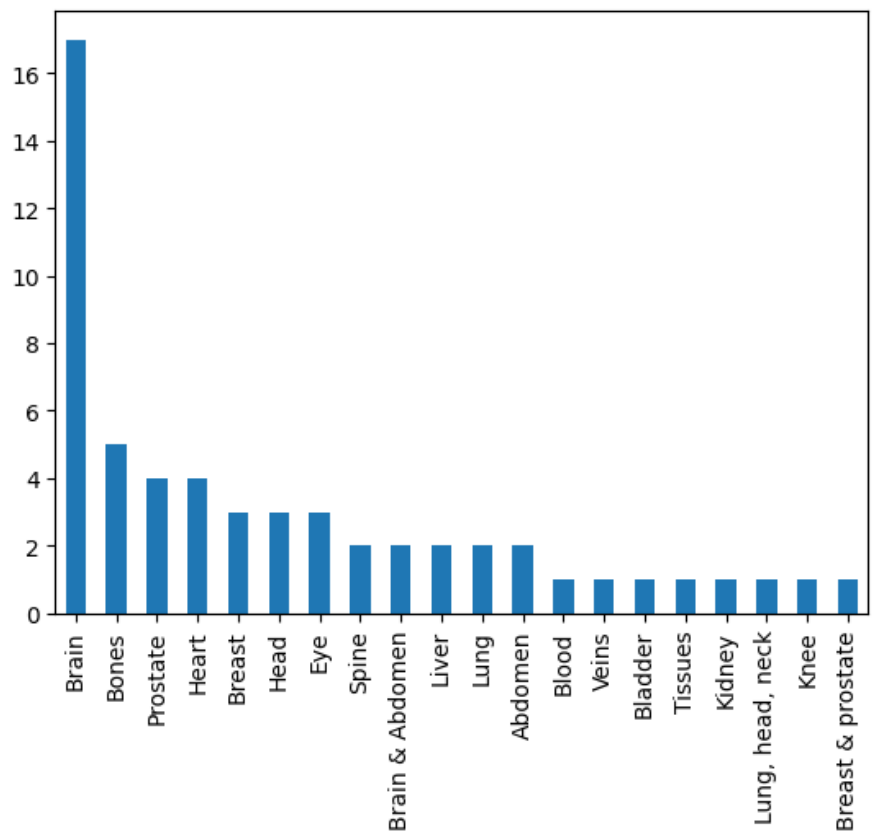

Figure 38. ROIs statistics